\documentclass[preprint, 12pt, 3p, authoryear]{elsarticle}
\makeatletter
\def\ps@pprintTitle{%
 \let\@oddhead\@empty
 \let\@evenhead\@empty
 \def\@oddfoot{}%
 \let\@evenfoot\@oddfoot}
\makeatother

\usepackage{tikz}
\usepackage{longtable,ragged2e,booktabs,siunitx,multirow,lipsum,array, threeparttablex}
\usepackage{natbib}
\usepackage{amssymb}
\usepackage{amsmath}
\usepackage{subcaption}
\usepackage[shortlabels]{enumitem}
\usepackage{mwe}
\usepackage{mathtools}
\usepackage{multicol}
\usepackage{graphicx}
\usepackage{color}
\usepackage[hang, flushmargin, multiple]{footmisc}
\usepackage[hyperfootnotes=false,colorlinks=true]{hyperref}
\usepackage{amsfonts}
\usepackage{graphicx}
\usepackage[utf8]{inputenc}
\usepackage[all]{xy}
\usepackage{csquotes}
\usepackage{ulem}
\usepackage{amsmath}
\usepackage{amsthm}
\usepackage{comment}
\usepackage{bbm}
\usepackage{arydshln}
\usepackage{setspace}
\usepackage[resetlabels]{multibib}

\newtheorem{remark}{Remark}

\newcolumntype{P}[1]{>{\RaggedRight\arraybackslash\hspace{0pt}}p{#1}}

\newcites{S}{References for Supplemental Materials}

\usetikzlibrary{shapes,decorations,arrows,calc,arrows.meta,fit,positioning}
\tikzset{
    -Latex,auto,node distance =1 cm and 1 cm,semithick,
    state/.style ={ellipse, draw, minimum width = 0.7 cm},
    point/.style = {circle, draw, inner sep=0.04cm,fill,node contents={}},
    bidirected/.style={Latex-Latex,dashed},
    el/.style = {inner sep=2pt, align=left, sloped}
}

\newpage

\pagebreak

\begin{document}

\begin{frontmatter}

\title{Grace periods in comparative effectiveness studies of sustained treatments}

\author[1,2]{Kerollos Nashat Wanis \corref{cor1}}
\author[3]{Aaron L. Sarvet}
\author[4]{Lan Wen}
\author[5]{Jason P. Block}
\author[5]{Sheryl L. Rifas-Shiman}
\author[1,6]{James M. Robins}
\author[1,5]{Jessica G. Young}

\address[1]{Department of Epidemiology, Harvard T. H. Chan School of Public Health, Boston, Massachusetts, USA}
\address[2]{Department of Breast Surgical Oncology, MD Anderson Cancer Center, Houston, TX, USA}
\address[3]{Department of Mathematics, École Polytechnique Fédérale de Lausanne, Switzerland}
\address[4]{Department of Statistics and Actuarial Science, University of Waterloo, Waterloo, Ontario, Canada}
\address[5]{Department of Population Medicine, Harvard Medical School and Harvard Pilgrim Health Care Institute, Boston, Massachusetts, USA}
\address[6]{Department of Biostatistics, Harvard T.H. Chan School of Public Health, Boston, Massachusetts, USA}
\cortext[cor1]{\textbf{Contact information for corresponding author:}\\
Kerollos N. Wanis, Department of Breast Surgical Oncology, MD Anderson Cancer Center, 1400 Pressler St, Unit 1434|FCT7.6088, Houston, TX, USA. \url{knwanis@gmail.com}}

\begin{abstract}
\begin{singlespace}
Researchers are often interested in estimating the effect of sustained use of a treatment on a health outcome. However, adherence to strict treatment protocols can be challenging for individuals in practice and, when non-adherence is expected, estimates of the effect of sustained use may not be useful for decision making. As an alternative, more relaxed treatment protocols which allow for periods of time off treatment (i.e., grace periods) have been considered in pragmatic randomized trials and observational studies. In this paper we consider the interpretation, identification, and estimation of treatment strategies which include grace periods. We contrast \textit{natural} grace period strategies which allow individuals the flexibility to take treatment as they would naturally do, with \textit{stochastic} grace period strategies in which the investigator specifies the distribution of treatment utilization. We estimate the effect of initiation of a thiazide diuretic or an angiotensin-converting enzyme inhibitor in hypertensive individuals under various strategies which include grace periods. %We find that the risk of adverse cardiovascular and cerebrovascular outcomes is higher under treatment with ACEIs, and that this difference is larger under grace period strategies which impose lower drug utilization.

\vspace{10pt}

%\noindent%
%{\it Keywords:} Causal Inference, Comparative Effectiveness Research, Hypertension, Lifetime and Survival Analysis

%\vspace{5pt}

%\noindent%
\noindent {\it Running head:} Grace periods for comparative effectiveness
\end{singlespace}
\end{abstract}
\end{frontmatter}

\section{Introduction}
\label{sec: Introduction}

Many randomized and observational studies aim to compare the safety and/or efficacy of different treatments indicated for the same disease diagnosis. These \textit{comparative effectiveness studies} have formed the basis for various pharmacologic recommendations in medical guidelines. American Heart Association (AHA) guidelines are one example. The AHA equivalently support either thiazide diuretics or angiotensin-converting enzyme inhibitors (ACEIs) as first-line treatment for hypertension \citep{whelton20182017}. A Cochrane review mainly based on randomized trials prior to the year 2000 informs the AHA recommendations, but acknowledges that evidence for ACEI effectiveness is of low quality \citep{musini2017pharmacotherapy}. A recent large scale observational study comparing initiation of various antihypertensive agents found that thiazides have better primary effectiveness than ACEIs \citep{suchard2019comprehensive}.

In comparative effectiveness studies, investigators might express interest in causal effects of sustained use, e.g., the effect on cardiovascular outcomes of taking a thiazide diuretic medication (versus an ACEI medication) continuously over the follow-up. However, it can be unrealistic to expect study participants to adhere to a strict treatment strategy (also termed a protocol or a regime) that demands continuous utilization of treatment. In causal inference, such strict protocols are examples of time-varying static deterministic treatment strategies: treatment rules requiring the same value of treatment for all individuals at a given time during the follow-up  \citep{hernan2020causal, robins2009estimation}. Deterministic strategies can be made more realistic by allowing cessation of treatment when unmanageable treatment-related toxicities occur \citep{hernan2017per,robins2000correcting}.

Adherence to strict deterministic rules is often challenging for individuals in practice, with adherent individuals often representing a small proportion of the study population. Interruptions in utilization of even greater than 60 days have been observed in one-third to one-half of individuals who initiate antihypertensives, with interruptions being more common in those using thiazides \citep{elliott2007persistence}. This can translate to (near) positivity violations \citep{petersen2012diagnosing, kennedy2019nonparametric, robins1986new}, which can preclude consistent estimation of effects of continuous utilization. Arguably, these positivity concerns also reflect the limited real-world clinical relevance of strict deterministic treatment strategies. As a pragmatic alternative, more relaxed study protocols have been defined in randomized trials to better reflect actual clinical practice \citep{sox2016pragmatic, loudon2015precis}, allowing for delays in treatment initiation or for periods of time off treatment after initiation, even in the absence of toxicities or other severe clinical events \citep{ford2016pragmatic}. 

These allowances can be put into practice using \textit{grace periods} - periods of time in which breaks from sustained use of treatment are permitted under the protocol. Treatment protocols in pragmatic trials allowing grace periods typically specify that individuals take treatment as they would naturally until the end of the grace period, at which time they must take treatment if they have not already done so. Causal effects of such \textit{natural grace period strategies} have also been considered in observational studies, typically with a single grace period prior to treatment initiation \citep{young2011comparative, cain2010start, huitfeldt2015methods, hernan2016using}.

However, the relative effectiveness of two medications is difficult to interpret under natural grace period strategies. Even if the medications have identical `pharmacological' efficacy, their effectiveness under natural grace period strategies may differ substantially if natural utilization of the drugs differs during grace periods.

Motivated by a study using data from Kaiser Permanente Colorado (KPC) comparing antihypertensive monotherapies, we will elaborate on the interpretation of effects under treatment strategies which include grace periods. As a compromise between strict protocols requiring continuous treatment utilization and natural grace period protocols, we describe alternative grace period protocols which are a flexible set of treatment strategies that can mitigate the highlighted issues inherent in the aforementioned extremes. We refer to this class of rules as \textit{stochastic} grace period strategies. We consider assumptions needed for identification of risk under this class of rules and how to reason about them under subject matter knowledge using Single World Intervention Graphs (SWIGs) \citep{richardson2013single}. We implement an estimator based on the efficient influence function, which permits the use of flexible machine learning algorithms for confounding adjustment.

\section{Observed data}\label{obsdata}

We consider a longitudinal study in which time-varying measurements are available for a random sample of $N$ individuals from a target population who are indicated to initiate one of two antihypertensives. Let $Z$ denote the medication initiated, with $Z=1$ denoting initiation of a thiazide diuretic and $Z=0$ an ACEI. To avoid distracting from central issues, we assume no loss to follow-up until Supplementary Materials Section~\ref{sec: extension to censoring}, and, although we consider a survival outcome in this manuscript, the results presented here can be generalized to settings with other types of outcomes. 

After treatment initiation, suppose the following naturally-arising covariates are measured in discrete time intervals (e.g., days, months), $k=1,\dots, K$, with $k=1$, the interval of treatment initiation (baseline), and $k=K$, the end of follow-up of interest (e.g., 36 months): a vector of covariates describing the individual's status in interval $k$, $L_{k}$ (e.g., new disease diagnoses); an indicator of adherence to the initiated medication in interval $k$, $A_{k}$; and an indicator of failure (e.g., adverse cardiovascular events) by the end of the interval, $Y_{k}$. $L_1$ may also include time-fixed characteristics (e.g., age at baseline, sex, history of diabetes mellitus at baseline).

In each interval, we assume the temporal ordering $(L_{k},A_{k},Y_{k})$. Overlines are used to represent an individual's history of covariates, treatment, or outcome from interval $k$ (e.g., $\overline{L}_{k}$). Individuals are outcome-free at the beginning of the study ($Y_{0}=0$), and, by notational convention, $L_{0}$ and $A_{0}$ are set to an arbitrary constant (e.g., $0$). We use lower-case letters to denote possible realizations of random variables (e.g., $a_k$ is a realization of $A_k$).

\section{Defining causal effects of treatment strategies with grace periods}
\label{sec:defining_grace_periods}

Define $Y_{k}^{z,g_m}$ as an individual's failure status by interval $k$ had, possibly contrary to fact, they followed a treatment strategy that ensures they initiate medication $z$ and then subsequently follow a particular time-varying grace period strategy $g_m$ that allows $m$ consecutive intervals without adhering to that medication. We can, in turn, quantify the comparative effectiveness of such strategies comparing different levels of $z$ on the outcome risk by $k$ as the counterfactual risk contrast
\begin{equation}
    \Pr[Y^{1,g_m}_{k}=1]\mbox { vs. } \Pr[Y^{0,g_m}_{k}=1].\label{effect}
\end{equation} 

In the special case of $m=0$ (no grace period), \eqref{effect} captures the effect of continuous utilization of the different medications, i.e., $Y^{z,g_0}_{k} = Y^{z,\overline{a}_K = 1}_{k}$. As such, static treatment regimes may always be understood as a special case of a grace-period regime. However, when $m>0$, the effect \eqref{effect}, and the absolute risk $\Pr[Y^{z,g_m}_{k}=1]$, may have any number of interpretations depending on how the treatment rules defining $g_m$ are chosen.

Following \citet{richardson2013single}, let $A_k^{z,g_m}$ denote the \textsl{natural} value of treatment at $k$ under a grace period strategy indexed by $z$ and $m$, and let $A_k^{z,g_m+}$ denote the \textit{intervention} value of treatment. $A_k^{z,g_m}$ is the treatment level that we would observe if that strategy were implemented up to $k$ but no intervention was made at $k$, while $A_k^{z,g_m+}$ is the treatment level that would be assigned at $k$ under the treatment strategy.

To facilitate definition of grace period strategies, define $R_k^{z,g_m+}$ as the number of consecutive intervals of follow-up prior to $k$ that an individual does not adhere to treatment under a grace period strategy indexed by $z,g_m$. Initializing, for $k=1$, $R^{z,g_m+}_1=0$, we recursively define 
\begin{equation}
   R_{k}^{z,g_m+} \coloneqq (R_{k-1}^{z,g_m+} + 1)\times (1-A_{k-1}^{z,g_m+}) \label{Rkdef}
\end{equation} $k=2,\ldots,K+1$. Therefore, $R_k^{z,g_m+}$ is a deterministic function of the intervention treatment history prior to $k$, counting consecutive intervals without treatment adherence and resetting to 0 whenever treatment is assigned in the prior interval. 

\textsl{Natural} grace period strategies have been the implicit type of grace period strategy in protocols of pragmatic trials as well as in the emulation of such trials with observational data \citep{cain2010start,young2011comparative}. We will now formally define these strategies, contrasting them with a family of alternative strategies, \textsl{stochastic} grace period strategies. The different nature of these definitions will have implications for interpretation of $\Pr[Y^{z,g_m}_{k}=1]$ and the effect \eqref{effect} but also for the assumptions needed for identification of these counterfactual quantities with observed data. We will discuss these assumptions in Section~\ref{sec:identification}. 

\subsection{Natural grace period strategies}

Under a natural grace period strategy, for individuals surviving to $k$ (i.e., $Y_{k-1}^{z,g_m}=0$), we define
\begin{align}
    A_k^{z,g_m+} \coloneqq \begin{cases}
     A_k^{z,g_m} &: R_k^{z,g_m+} < m, \\
     1 &: R_k^{z,g_m+} = m. \\
    \end{cases}\label{natural}
\end{align}

We emphasize the distinctive feature of natural grace period regimes: that $A_k^{z,g_m+}~:=~A_k^{z,g_m}$ whenever $R_k^{z,g_m+}~<~m$. In words, surviving individuals will take treatment as they would \textit{naturally} (no intervention will be made) at times $k$ that are during a grace period. Because intervention treatment values are equal to natural treatment values at all times during a grace period (when $R^{z,g_m+}_k<m$) under this rule, it follows by the definition of $R_k^{z,g_m+}$ in \eqref{Rkdef} that treatment assignment at the end of a grace period (when $R_k^{z,g_m+} = m$) is a function of the natural treatment history. Therefore, a natural grace period strategy is an intervention that depends on the natural treatment history. For examples of other interventions that depend on the natural treatment history, see \citet{richardson2013single, young2014identification, haneuse2013estimation, diaz2021nonparametric}. 

\begin{remark}
\begin{singlespace}
In the special case of $m > K$, for a natural grace period strategy, \eqref{effect} is equivalent to the effect of initiating medication then making no intervention on adherence at any time point (an analogue of the `intention-to-treat' effect) since $R_k^{z,g_m+} < m$ at all times $k$.
\end{singlespace}
\end{remark}

\subsection{Stochastic grace period strategies}
In contrast to a natural grace period strategy, consider an alternative class of strategies indexed by $z,g_m$ which we will generally define as follows: for those surviving under the strategy to $k$ (i.e., $Y_{k-1}^{z,g_m}=0$), let $X_k\sim U(0,1)$ be an exogenous randomly drawn value from a standard uniform distribution and, using an \textit{investigator}-specified function $\pi^{inv}_k$, define  
\begin{align}
    A_k^{z,g_m+} \coloneqq \begin{cases}
     1 &: R_k^{z,g_m+} < m \mbox{ and } X_k \leq \pi^{inv}_k(\overline{L}_k^{z,g_m},\overline{A}_{k-1}^{z,g_m+},z), \\
     0 &: R_k^{z,g_m+} < m \mbox{ and } X_k>\pi^{inv}_k(\overline{L}_k^{z,g_m},\overline{A}_{k-1}^{z,g_m+},z), \\
     1 &: R_k^{z,g_m+} = m.
    \end{cases}\label{stochastic}
\end{align}
where $\overline{L}_k^{z,g_m}$ is the history of the values of the covariates at $k$ under the strategy and $\pi^{inv}_k$ gives the probability of receiving treatment during a grace period, which may depend on the covariate and intervention treatment history $(\overline{L}_k^{z,g_m},\overline{A}_{k-1}^{z,g_m+})$, as well as the medication initiated, $z$, under the strategy. When $\pi^{inv}_k$ depends trivially on some of its arguments, we will omit those arguments from the function (e.g., as in Supplementary Materials Section~\ref{section: stochastic grace period examples}). We will refer to any strategy within the class \eqref{stochastic} as a \textsl{stochastic grace period strategy}. In words, under a stochastic grace period strategy, an intervention on treatment will occur at all times $k$ that are during a grace period -- individuals will either be forced to take treatment or not take treatment depending on the random draw of the exogenous random variable $X_k$ and the investigator chosen treatment probability $\pi^{inv}_k$. The choice of $\pi^{inv}_k$ allows investigators to control the amount and distribution of non-adherence allowed under the grace period treatment strategy. Investigators might choose to induce a discrete analogue of a particular continuous distribution, as we illustrate in Supplementary Materials Section~\ref{section:constructing_stochastic}. Alternatively, a constant $\pi^{inv}_k$ may be a parsimonious choice. 

We can see from the definition \eqref{stochastic} that, unlike a natural grace period strategy, a stochastic grace period strategy does \textsl{not} depend on the natural treatment history but, at most, is allowed to depend on the intervention treatment history, the covariate history, the initiated medication, and an exogenous randomizer.  Recall, by definition \eqref{Rkdef}, that $R_k^{z,g_m+}$ is a function of the intervention treatment history $\overline{A}_{k-1}^{z,g_m+}$. Thus, an investigator might restrict the dependence of the stochastic grace period strategies as defined in \eqref{stochastic} to $R_k^{z,g_m+}$ rather than the entire intervention treatment history. 

A stochastic grace period strategy with $m=3$ following initiation of an ACEI could be implemented by having investigators flip a fair coin for each individual at the beginning of each month. If the result of the coin flip is heads, then the individual receives a month's supply of ACEI. If the result is tails, then the individual does not receive medication. However, if an individual has gone 3 consecutive months without medication (due to prior coin flips), then no coin flip is necessary because they must receive a dispensation. This stochastic grace period strategy depends only on $R_k^{z,g_m+}$, and sets $\pi^{inv}_k=0.5$. By contrast, a natural grace period strategy with $m=3$ would allow individuals to miss their monthly ACEI dispensations over the course of follow-up according to their natural inclinations so long as they never miss more than 3 in a row, at which point they would be required to receive a dispensation. The distribution of treatment adherence is therefore not under the control of the investigator when a natural grace period strategy is chosen. In Supplementary Materials Section~\ref{section: stochastic grace period examples} and \ref{section: grace period examples} we consider other specific examples of grace period strategies.

\section{Implications of the choice of grace period strategy for interpretation in comparative effectiveness research}
\label{section: written_example}

To consider the implications of a particular choice of grace period strategy on interpretation of the counterfactual absolute risk $\Pr[Y^{z,g_m}_{k}=1]$ and the effect \eqref{effect}, we will consider our motivating study comparing antihypertensive agents under grace period strategies. Suppose that individuals initiate a medication from the ACEI class ($z=0$) or a medication from the thiazide diuretic class ($z=1$) and we compare expected potential outcomes, as in \eqref{effect}, under strategies that allow grace periods of $m=3$ months.

First, consider natural grace period strategies \eqref{natural}. Under these strategies, the chance of receiving a pharmacy dispensation may vary for each individual in any particular month within a grace period. Suppose further that, while there indeed is some non-adherence to the initiated medication, all individuals \textsl{adhere to the natural grace period protocols} in that no individual exceeds 3 consecutive months without dispensing their initiated medication. It might appear that the comparison of thiazide diuretics and ACEIs under their natural grace period protocols is a fair comparison because the rules for receiving pharmacy dispensations are the same in name under both natural strategies; they are both grace period strategies with the same choice of $m$. However, because pharmacy dispensing depends on the natural adherence process when a natural grace period strategy is chosen, the distribution of dispensing may differ considerably in the two arms if adherence depends on $z$ in the target population. 

As an extreme example, consider a scenario in which individuals who initiate an ACEI naturally take treatment on the first interval of every grace period with certainty, whereas individuals who intiate a thiazide naturally never take treatment until the last interval of every grace period. In other words, those with $Z=0$ naturally take treatment in every interval, whereas those with $Z=1$ only take treatment every $m^{th}$ ($3^{rd}$) interval. In such a setting, individuals with $Z=0$ and following a natural grace period strategy will have dispensed far more often over the course of follow-up than those with $Z=1$. 

In contrast, stochastic grace period strategies allow investigators to overcome this issue by contrasting regimes that fix adherence to an investigator-specified process, which can be equivalent for both initiated medications when $\pi^{inv}_k$ does not depend on $z$ or any covariates affected by $z$. For example, consider stochastic grace period strategies \eqref{stochastic} with \mbox{$\pi^{inv}_k(z=0)=\pi^{inv}_k(z=1)=0.5$} for all $k$. In words, under these strategies, the chance of receiving a pharmacy dispensation is 50\% for every individual and for any month within a grace period regardless of initiated antihypertensive. 

Because natural grace period strategies depend on the natural adherence process, if either antihypertensive has a non-null effect on adverse cardiovascular outcomes compared to taking no treatment at all, then $\Pr[Y^{0,g_m}_{k}=1]$ may be very different from $\Pr[Y^{1,g_m}_{k}=1]$ when $g_m$ represents a natural grace period strategy even if they would be identical under any stochastic grace period strategy that sets $\pi^{inv}_k(\overline{a}_k,z=0) = \pi^{inv}_k(\overline{a}_k,z=1)$ for all values $\overline{a}_k$ and for all $k$. Natural grace period strategies therefore present challenges to investigators seeking specific conclusions about the `pharmacological' effectiveness of ACEIs compared to thiazide diuretics because the counterfactual risk contrast depends not only on the effectiveness of the two medications but also on their natural adherence processes in the study population. Because natural adherence processes are not specified by investigators and, unlike in our example, will generally be unknown, providing an unambiguous interpretation of a counterfactual risk contrast between natural grace period strategies is difficult. 

In Supplementary Materials Section~\ref{sec: simulation example} we present results from a simulation example that contrasts estimates from natural grace period strategies and simple stochastic grace period strategies. 

\section{Exchangeability and the choice of grace period strategy}
\label{sec:identification}

We now consider the nature of exchangeability (alternatively, `no unmeasured confounding') conditions required for identification of the counterfactual absolute risk $\Pr[Y^{z,g_m}_{k}=1]$ for a particular choice of grace period strategy using the observed data of the type described in Section \ref{obsdata}. In considering these conditions, we rely on results from \citet{richardson2013single} to make explicit that, not only is the interpretation of $\Pr[Y^{z,g_m}_{k}=1]$ and corresponding causal effects dependent on how the investigator chooses to define the strategy during the grace period, but the strength of exchangeability conditions required for identifying  $\Pr[Y^{z,g_m}_{k}=1]$ with the study data will also depend on this choice. Throughout, we will assume that consistency, which allows linking counterfactual to factual random variables, holds.

We use Single World Intervention Graphs (SWIGs) to represent the assumptions about the underlying process that produces outcomes under a grace period strategy \citep{richardson2013single}. SWIGS, unlike more traditional causal directed acyclic graphs (DAGs), explicitly depict counterfactual variables which makes them better suited for reasoning about exchangeability conditions, which directly concern counterfactual independencies \citep{shpitser2020multivariate, sarvet2020graphical}. SWIGs are particularly illuminating in the case of treatment strategies that depend on natural treatment values, such as the natural grace period treatment strategies we consider here, allowing explicit depiction of both natural (e.g., $A_k^{z,g_m}$) and intervention (e.g., $A_k^{z,g_m+}$) treatment values. The construction of SWIGs is described by \citet{richardson2013singleprim} and is reviewed in Supplementary Materials Section~\ref{section: swig construction}.

Exchangeability for $\Pr[Y^{z,g_m}_{k}=1]$ can be read from a SWIG that shows, for each $j=1,\ldots, k$, \textit{no unblocked backdoor paths} connecting the outcome $Y^{z,g_m}_k$ and the natural treatment value $A_j^{z,g_m}$ conditional on past natural values of the measured covariates and treatment, past survival, as well as past intervention treatment values, under the strategy $z,g_m$ \citep[Lemma 33]{richardson2013single}. We will subsequently refer to this as the backdoor criterion. An unblocked backdoor path between two nodes $A_j^{z,g_m}$ and $Y^{z,g_m}_k$ is an unblocked path that connects the nodes and that emanates from an arrow \textit{into} $A_j^{z,g_m}$ (i.e., $Y^{z,g_m}_k \ldots \rightarrow A_j^{z,g_m}$). An unblocked path that connects $A_j^{z,g_m}$ and $Y^{z,g_m}_k$, but that emanates from an arrow \textit{out} of $A_j^{z,g_m}$ (i.e., $A_j^{z,g_m} \dashrightarrow \ldots Y^{z,g_m}_k$), is therefore not a \textit{backdoor} path. Additionally, we will require the more familiar assumption that there are no unblocked backdoor paths between $Z$ and $Y^{z,g_m}_k$ conditional on $L_1$ which is expected to hold in a trial where $Z$ is randomized, possibly within levels of $L_1$. 
 
When treatment has an effect on survival, exchangeability will not only be violated in the presence of unmeasured common causes of treatment and survival, but also (i) in the presence of unmeasured common causes of treatment and future covariates when treatment assignment depends on those covariates under the specified regime, and/or (ii) in the presence of unmeasured common causes of any two treatment variables when treatment assignment depends on past natural treatment values as it does in the natural grace period regimes \eqref{natural}. We now illustrate this with examples.
 
\subsection{Causal graph examples}

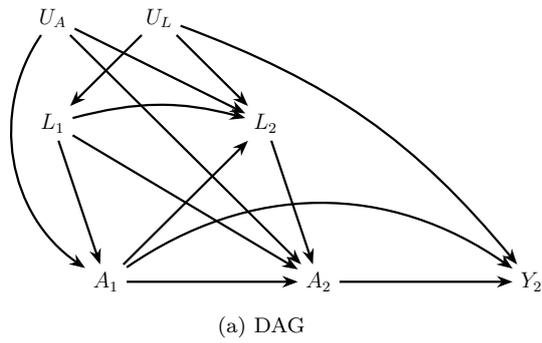
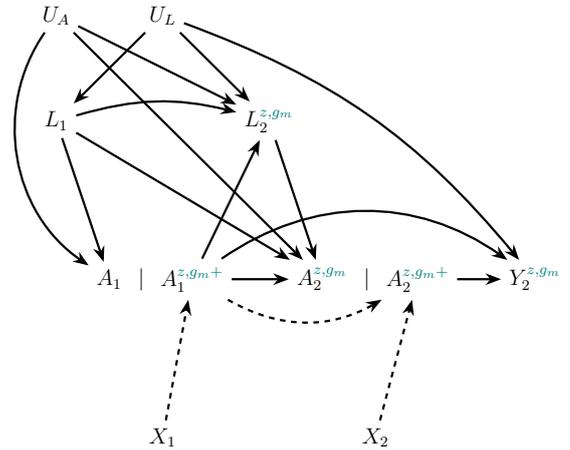
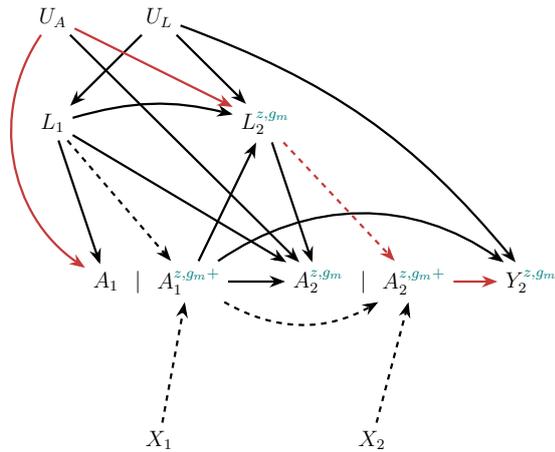
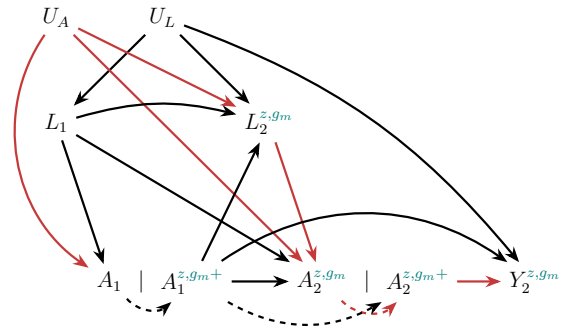
\begin{figure}
    \centering
\begin{subfigure}[b]{0.45\textwidth}
\scalebox{0.7}{
\begin{tikzpicture}
\begin{scope}[every node/.style={thick,draw=none}]
    
    \node (UA) at (0, 5) {$U_A$};
    \node (UL) at (2, 5) {$U_L$};
    
    \node (L1)   at ( 0, 3 ) {$L_1$};
    \node (L2)  at ( 4, 3 ) {$L_2$};

    \node (A1)   at ( 1, 0 ) {$A_1$};
    \node (A2)  at ( 5, 0 ){$A_2$};

    \node (Y)   at ( 9, 0 ) {$Y_{2}$};
\end{scope}
\begin{scope}[>={Stealth[]},
              every node/.style={fill=white,circle},
              every edge/.style={draw=black,very thick}]
    
    \path [->] (UL) edge (L1);
    \path [->] (UA) edge (L2);
    \path [->] (UA)[bend right=45] edge (A1);
    \path [->] (UA) edge (A2);
    \path [->] (UL) edge (L2);
    \path [->] (UL)[bend left=15] edge (Y);
    
    \path [->] (L1)  edge (A1);
    \path [->] (L1)  edge (A2);
    
    \path [->] (L2)  edge (A2);
    
    \path [->] (L1)[bend left=15]  edge (L2);
    
    \path [->] (A1)  edge (L2);
    
    \path [->] (A1)  edge (A2);
    
    \path [->] (A1)[bend left=35]  edge (Y);
    \path [->] (A2)  edge (Y);
\end{scope}
\end{tikzpicture}
}
\caption{DAG}
\end{subfigure}
  \hfill
\begin{subfigure}[b]{0.45\textwidth}
\scalebox{0.7}{
\begin{tikzpicture}
\begin{scope}[every node/.style={thick,draw=none}]

    \node (UA) at (0, 5) {$U_A$};
    \node (UL) at (2, 5) {$U_L$};

    \node (L1)   at ( 0, 3 ) {$L_1$};
    \node (L2)  at ( 4, 3 ) {$L_2 ^{\textcolor{teal}{z,g_m}}$};

    \node (A1)   at ( 1, 0 ) {$A_1$};
        \node (midA1)[right = 0 of A1,    anchor=west] {$\mid$};
        \node (A1+)  [right = 0 of midA1, anchor=west] {$A_1 ^{\textcolor{teal}{z,g_m+}}$};
    \node (A2)  at ( 5, 0 ){$A_2^{\textcolor{teal}{z,g_m}}$};
        \node (midA2)[right = 0 of A2,    anchor=west] {$\mid$};
        \node (A2+)  [right = 0 of midA2, anchor=west] {$A_2 ^{\textcolor{teal}{z,g_m+}}$}; 
    \node (X1)  at ( 2, -3 ) {$X_1$};
    \node (X2) at ( 6, -3 ) {$X_2$};

    \node (Y)   at ( 9, 0 ) {$Y_{2}^{\textcolor{teal}{z,g_m}}$};
\end{scope}
\begin{scope}[>={Stealth[]},
              every node/.style={fill=white,circle},
              every edge/.style={draw=black,very thick}]
    
    \path [->] (UL) edge (L1);
    \path [->] (UA) edge (L2);
    \path [->] (UA)[bend right=45] edge (A1);
    \path [->] (UA) edge (A2);
    \path [->] (UL) edge (L2);
    \path [->] (UL)[bend left=15] edge (Y);
    
    \path [->] (L1)  edge (A1);
    \path [->] (L1)  edge (A2);
    
    \path [->] (L2)  edge (A2);
    
    \path [->] (L1)[bend left=15]  edge (L2);
    
    \path [->] (X1)[dashed]  edge (A1+);
    \path [->] (X2)[dashed]  edge (A2+);

    \path [->] (A1+)[bend right=30, dashed]  edge (A2+);
    
    \path [->] (A1+)  edge (L2);
    
    \path [->] (A1+)  edge (A2);
    
    \path [->] (A1+)[bend left=35]  edge (Y);
    \path [->] (A2+)  edge (Y);
\end{scope}
\end{tikzpicture}
}
\caption{SWIG for a stochastic grace period strategy that only depends on prior assigned treatment}
\end{subfigure}
\vfill
\begin{subfigure}[b]{0.45\textwidth}
\scalebox{0.7}{
\begin{tikzpicture}
\begin{scope}[every node/.style={thick,draw=none}]

    \node (UA) at (0, 5) {$U_A$};
    \node (UL) at (2, 5) {$U_L$};

    \node (L1)   at ( 0, 3 ) {$L_1$};
    \node (L2)  at ( 4, 3 ) {$L_2 ^{\textcolor{teal}{z,g_m}}$};

    \node (A1)   at ( 1, 0 ) {$A_1$};
        \node (midA1)[right = 0 of A1,    anchor=west] {$\mid$};
        \node (A1+)  [right = 0 of midA1, anchor=west] {$A_1 ^{\textcolor{teal}{z,g_m+}}$};
    \node (A2)  at ( 5, 0 ){$A_2^{\textcolor{teal}{z,g_m}}$};
        \node (midA2)[right = 0 of A2,    anchor=west] {$\mid$};
        \node (A2+)  [right = 0 of midA2, anchor=west] {$A_2 ^{\textcolor{teal}{z,g_m+}}$}; 
    \node (X1)  at ( 2, -3 ) {$X_1$};
    \node (X2) at ( 6, -3 ) {$X_2$};

    \node (Y)   at ( 9, 0 ) {$Y_{2}^{\textcolor{teal}{z,g_m}}$};
\end{scope}
\begin{scope}[>={Stealth[]},
              every node/.style={fill=white,circle},
              every edge/.style={draw=black,very thick}]
    
    \path [->] (UL) edge (L1);
    \path [->] (UA) edge[color=gray!45!red] (L2);
    \path [->] (UA)[bend right=45] edge[color=gray!45!red] (A1);
    \path [->] (UA) edge (A2);
    \path [->] (UL) edge (L2);
    \path [->] (UL)[bend left=15] edge (Y);
    
    \path [->] (L1)  edge (A1);
    \path [->] (L1)  edge (A2);
    
    \path [->] (L2)  edge (A2);
    
    \path [->] (L1)[dashed]  edge (A1+);
    \path [->] (L2)[dashed]  edge[color=gray!45!red] (A2+);
    
    \path [->] (L1)[bend left=15]  edge (L2);
    
    \path [->] (X1)[dashed]  edge (A1+);
    \path [->] (X2)[dashed]  edge (A2+);

    \path [->] (A1+)[bend right=30, dashed]  edge (A2+);
    
    \path [->] (A1+)  edge (L2);
    
    \path [->] (A1+)  edge (A2);
    
    \path [->] (A1+)[bend left=35]  edge (Y);
    \path [->] (A2+)  edge[color=gray!45!red] (Y);
\end{scope}
\end{tikzpicture}
}
\caption{SWIG for a stochastic grace period strategy that depends on prior assigned treatment and measured covariates}
\end{subfigure}
\hfill
\begin{subfigure}[b]{0.45\textwidth}
\scalebox{0.7}{
\begin{tikzpicture}
\begin{scope}[every node/.style={thick,draw=none}]
    
    \node (UA) at (0, 5) {$U_A$};
    \node (UL) at (2, 5) {$U_L$};

    \node (L1)   at ( 0, 3 ) {$L_1$};
    \node (L2)  at ( 4, 3 ) {$L_2 ^{\textcolor{teal}{z,g_m}}$};

    \node (A1)   at ( 1, 0 ) {$A_1$};
        \node (midA1)[right = 0 of A1,    anchor=west] {$\mid$};
        \node (A1+)  [right = 0 of midA1, anchor=west] {$A_1 ^{\textcolor{teal}{z,g_m+}}$};
    \node (A2)  at ( 5, 0 ){$A_2^{\textcolor{teal}{z,g_m}}$};
        \node (midA2)[right = 0 of A2,    anchor=west] {$\mid$};
        \node (A2+)  [right = 0 of midA2, anchor=west] {$A_2 ^{\textcolor{teal}{z,g_m+}}$}; 

    \node (Y)   at ( 9, 0 ) {$Y_{2}^{\textcolor{teal}{z,g_m}}$};
\end{scope}
\begin{scope}[>={Stealth[]},
              every node/.style={fill=white,circle},
              every edge/.style={draw=black,very thick}]

    \path [->] (UL) edge (L1);
    \path [->] (UA) edge[color=gray!45!red] (L2);
    \path [->] (UA)[bend right=45] edge[color=gray!45!red] (A1);
    \path [->] (UA) edge[color=gray!45!red] (A2);
    \path [->] (UL) edge (L2);
    \path [->] (UL)[bend left=15] edge (Y);

    \path [->] (L1)  edge (A1);
    \path [->] (L1)  edge (A2);

    \path [->] (L2)  edge[color=gray!45!red] (A2);
    
    \path [->] (L1)[bend left=15]  edge (L2);

    \path [->] (A1)[bend right=45, dashed]  edge (A1+);
    \path [->] (A2)[bend right=45, dashed]  edge[color=gray!45!red] (A2+);

    \path [->] (A1+)[bend right=30, dashed]  edge (A2+);

    \path [->] (A1+)  edge (L2);

    \path [->] (A1+)  edge (A2);
    
    \path [->] (A1+)[bend left=35]  edge (Y);
    \path [->] (A2+)  edge[color=gray!45!red] (Y);
\end{scope}
\end{tikzpicture}
}
\caption{SWIG for a natural grace period strategy}
\end{subfigure}
\caption{Causal graphs with $U_L$ and $U_A$ representing unmeasured variables. SWIGs in (b)-(d) each represent a different grace period strategy specific transformation of the causal DAG in (a), with (b) representing a stochastic grace period strategy depending only on assigned treatment history, (c) representing a stochastic grace period strategy additionally depending on measured covariates, and (d) representing a natural grace period strategy. For simplicity, the node $Z$, an indicator of the medication initiated at baseline, and $Y_1^{z,g_m}$ are not shown. Red paths violate the backdoor criterion.}
\label{fig: Assumption 1a holds}
\end{figure}

Figure~\ref{fig: Assumption 1a holds}a is a causal DAG representing a data generating mechanism where $U_A$ and $U_L$ are unmeasured. Figures~\ref{fig: Assumption 1a holds}b-\ref{fig: Assumption 1a holds}d depict SWIGs specific to different grace period strategies: a stochastic grace period strategy depending only on $\overline{A}_k^{z,g_m+}$ (Figure~\ref{fig: Assumption 1a holds}b); a stochastic grace period strategy additionally depending on measured covariates $\overline{L}_k^{z,g_m}$ (Figure~\ref{fig: Assumption 1a holds}c); and a natural grace period strategy (Figure~\ref{fig: Assumption 1a holds}d). Exchangeability can be read from all three SWIGs by the backdoor path criterion defined above. 

We can see that this criterion holds in the SWIG representing a stochastic grace period strategy that only depends on past treatment (Figure~\ref{fig: Assumption 1a holds}b). However, the criterion fails in both Figures~\ref{fig: Assumption 1a holds}c and \ref{fig: Assumption 1a holds}d, which represent a stochastic grace period strategy that also depends on measured covariate history, and a natural grace period strategy, respectively. Removing the red arrow from $U_A$ into $L_{2}^{z, g_m}$ would suffice for the backdoor criterion to hold in Figure~\ref{fig: Assumption 1a holds}c, but not in Figure~\ref{fig: Assumption 1a holds}d. In Figure~\ref{fig: Assumption 1a holds}d, it is also necessary to remove at least one of the red arrows $U_A \rightarrow A_1$ or $U_A \rightarrow A_2^{z,g_m}$ for the backdoor criterion to hold. 

This illustrates how the presence of an unmeasured common cause of two treatments (here, adherence over time) leads to violations of exchangeability assumptions required for identification of natural grace period strategies, even when the unmeasured variable has no directed arrow into any variables other than treatments. Unmeasured variables with these properties are likely common in many settings, including our data example. Examples include distaste about features of an initiated medication, such as the size or taste of the drug, which could impact adherence but might not otherwise exert effects on the outcome (except through adherence). 

\section{The g-formula and positivity}
\label{sec: gformula}

Suppose the underlying observed data generating mechanism is such that exchangeability holds as defined in Section~\ref{sec:identification}. It follows that the counterfactual risk $\Pr[Y^{z,g_m}_{k}=1]$ may be written as Robins's (generalized) g-formula \citep{robins1986new}, 
\begin{align}
    \sum_{k=1}^{K}\sum_{\overline{a}_k}
    &\sum_{\overline{l}_k} \Pr[Y_{k} = 1 \mid \overline{A}_{k} = \overline{a}_{k}, \overline{L}_{k} = \overline{l}_{k}, Y_{k-1}=0, Z=z] \times \nonumber \\
    \prod_{j=1}^{k} \{&\Pr[Y_{j-1} = 0 \mid \overline{A}_{j-1} = \overline{a}_{j-1}, \overline{L}_{j-1} = \overline{l}_{j-1}, Y_{j-2}=0, Z=z] \times \label{gformula} \\
    &f(l_{j} \mid \overline{a}_{j-1}, \overline{l}_{j-1}, y_{j-1} = 0, z) \times \nonumber \\
    &f^{g_m}(a_j \mid  \overline{a}_{j-1}, \overline{l}_j, y_{j-1} = 0, z)\} \nonumber,
\end{align}
provided that this function is well-defined. In \eqref{gformula}, $\sum\limits_{\overline{a}_k}$ and $\sum\limits_{\overline{l}_k}$ denote sums over all possible levels of treatment and confounder history, respectively; $\Pr[Y_{k} = 1 \mid \overline{A}_{k} = \overline{a}_{k}, \overline{L}_{k} = \overline{l}_{k}, Y_{k-1}=0, Z=z]$ is the observed probability of failing in interval $k$ among those observed to survive through $k-1$ with observed treatment and measured confounder history equal to the particular level $(\overline{a}_{k}, \overline{l}_{k})$ and initiated medication $z$, and $f(l_{k} \mid \overline{a}_{k-1}, \overline{l}_{k-1}, y_{k-1} = 0, z)$ is the observed joint probability of having the particular level $l_k$ of the measured confounders in interval $k$ among those observed to survive through $k-1$ with observed treatment and measured confounder history equal to the particular level $(\overline{a}_{k-1}, \overline{l}_{k-1})$ and initiated medication $z$. For notational simplicity we write the g-formula as a sum conceptualizing all covariates as discrete with perhaps a very large number of possible levels.  Generalizations to allow continuous covariates would include replacing sums with integrals.

Unlike the other terms comprising (\ref{gformula}), $f^{g_m}(a_k \mid \overline{l}_k, \overline{a}_{k-1}, y_{k-1} = 0, z)$ does not generally denote an observed probability. Rather this quantifies, among individuals surviving through $k-1$ with observed confounder and treatment history $(\overline{a}_{k-1}, \overline{l}_{k-1})$ and initiated medication $z$, the counterfactual probability of receiving treatment level $a_k$ under the grace period strategy. This probability depends on the investigators' specification $z,g_m$. Because $A_k$ is binary, we have  
\begin{align}
\label{fintdef}
  & f^{g_m}(a_k \mid \overline{l}_k, \overline{a}_{k-1}, y_{k-1} = 0, z) = 
    \begin{cases}
    \pi^{g_m}_k(\overline{l}_k, \overline{a}_{k-1},z)  &\ : a_k = 1, \\
    1-\pi^{g_m}_k(\overline{l}_k, \overline{a}_{k-1},z) &\ : a_k = 0,
    \end{cases}
\end{align}
where $\pi^{g_m}_k(\overline{l}_k, \overline{a}_{k-1},z)$ is the chance of receiving treatment conditional on having measured confounder and treatment history level $(\overline{l}_k, \overline{a}_{k-1})$ and initiated medication $z$ associated with the strategy $z,g_m$.  

Following \eqref{natural}, under a natural grace period strategy, when the exchangeability condition in Section~\ref{sec:identification} holds, we have 
\begin{align}
    \label{eq: f_int_natural_definition}
        & \pi^{g_m}_k(\overline{l}_k, \overline{a}_{k-1},z) = 
            \begin{cases}
            \pi^{obs}_k(\overline{l}_k, \overline{a}_{k-1},z)  &\ : r_k < m, \\
             1 &\ : r_k = m,
            \end{cases}
\end{align}
where $\pi^{obs}_k(\overline{l}_k, \overline{a}_{k-1},z)$ is the \textsl{observed} propensity score at $k$ \citep{rosenbaum1983central}. In words, when the exchangeability condition discussed in Section~\ref{sec:identification} holds, then the intervention propensity score at $k$ under a natural grace period strategy will be equal to the observed propensity score at $k$ for any individual who is observed to be in a grace period at $k$. By contrast, the intervention propensity score associated with a natural grace period strategy will be 1 at $k$ for any individual who is observed to reach the end of a grace period at $k$. 

Alternatively, by \eqref{stochastic} under a stochastic grace period strategy we have
\begin{align}
    \label{eq: f_int_stochastic_definition}
        & \pi^{g_m}_k(\overline{l}_k, \overline{a}_{k-1},z) = 
            \begin{cases}
            \pi^{inv}_k(\overline{l}_k, \overline{a}_{k-1},z)  &\ : r_k < m, \\
             1 &\ : r_k = m.
            \end{cases}
\end{align}
In words, the intervention propensity score at $k$ associated with a stochastic grace period strategy evaluated at the particular history $(\overline{l}_k, \overline{a}_{k-1},z)$ will be equal to the investigator specified treatment probability $\pi^{inv}_k(\overline{l}_k, \overline{a}_{k-1},z)$ for any individual who is observed to be in a grace period at $k$. As in the case of a natural grace period strategy, the  intervention propensity score under a stochastic grace period strategy will be 1 at $k$ for any individual who is observed to reach the end of a grace period at $k$. 

\begin{remark}
\begin{singlespace}
When a stochastic grace period strategy is defined by the choice $\pi^{inv}_k(\overline{l}_k, \overline{a}_{k-1},z)=\pi^{obs}_k(\overline{l}_k, \overline{a}_{k-1},z)$, then the g-formula \eqref{gformula} will be identical for this stochastic grace period strategy and a natural grace period strategy with the same choice of $m$ as long as all necessary exchangeability conditions hold. If exchangeability conditions for natural grace period strategies do not hold, then this same g-formula may still identify risk under a stochastic grace period strategy with probability of treatment equal to the observed distribution of treatment (see arguments in Section~\ref{sec:identification}).
\end{singlespace}
\end{remark}

Given positivity for the initiation component of a strategy $z,g_m$ (that is, if $f(l_1)>0$ then $\Pr[Z=z|L_1=l_1]>0$) the following generalized positivity condition is sufficient to ensure that \eqref{gformula} is well-defined \citep{robins1986new,young2014identification}:  

 \begin{align}
&\mbox{if }f(\overline{l}_k,\overline{a}_{k-1}, y_{k-1}=0,z)>0\mbox{ then }
f^{g_m}(a_k \mid \overline{l}_k,\overline{a}_{k-1}, y_{k-1}=0, z) > 0\nonumber\\ &\implies f(a_k \mid \overline{l}_k, \overline{a}_{k-1}, y_{k-1}=0, z) > 0, k=1,\ldots,K\label{genpos},
 \end{align}
where $f(a_k \mid \overline{l}_k, \overline{a}_{k-1}, y_{k-1} = 0, z)$ is the observed conditional probability of receiving treatment at $k$. Analogous to the relation \eqref{fintdef}, because treatment receipt status at $k$ is binary, $f(a_k \mid \overline{l}_k, \overline{a}_{k-1}, y_{k-1} = 0, z)$ is fully determined by the observed propensity score at $k$, $\pi^{obs}_k(\overline{l}_k, \overline{a}_{k-1},z)$. In words, the generalized positivity condition \eqref{genpos} states that, at each $k=1,\ldots,K$, conditional on any covariate and treatment history that can arise in the study population under the strategy $z,g_m$, any treatment level $a_k$ that is possible under the strategy must also be possible to observe among such individuals under no intervention. Thus, like the exchangeability assumption, evaluation of whether positivity holds is dependent both on the observed data generating mechanism and on the choice of strategy $z,g_m$. For example, when $z,g_m$ is chosen as a natural grace period strategy, we can see by \eqref{eq: f_int_natural_definition} that positivity is guaranteed at $k$ for any observed history compatible with following a natural grace period strategy to interval $k$, and consistent with being in a grace period at interval $k$ ($r_k<m$), because $\pi^{g_m}_k(\overline{l}_k, \overline{a}_{k-1},z)=\pi^{obs}_k(\overline{l}_k, \overline{a}_{k-1},z)$. In words, positivity is guaranteed for natural grace period strategies, within grace periods, because the probability of assigned treatment is precisely equal to the probability in the observed data. However, this guarantee cannot generally be made for stochastic grace period strategies, except for some choices of $\pi^{inv}_k$ that may be functions of $\pi^{obs}_k$ \citep{kennedy2019nonparametric,sarvet2023longitudinal,wen2023intervention}, including when $\pi^{inv}_k(\overline{l}_k, \overline{a}_{k-1},z)=\pi^{obs}_k(\overline{l}_k, \overline{a}_{k-1},z)$. In this special case of stochastic grace period strategies, the probability of assigned treatment during grace periods is similarly identical to the probability in the observed data. In either case of stochastic or natural grace period strategies, positivity is not guaranteed at $k$ for any observed history consistent with $r_k=m$. 

\section{Estimation}
\label{sec: AIPW}

If all the terms in the g-formula \eqref{gformula} can be consistently estimated using parametric models, and the assumptions discussed in Sections~\ref{sec:identification} and \ref{sec: gformula} hold, then g-computation \citep{hernan2020causal,mcgrath2020gformula} will consistently estimate risk under the strategy $z,g_m$. Other methods which rely on correctly specified parametric models include estimators motivated by inverse probability weighted representations of \eqref{gformula}. We describe one such estimator in Supplementary Materials Section~\ref{sec: IPW}. 

However, as is generally the case, when investigators are not able to specify correct parametric models, combining data adaptive estimators with multiple robustness and sample splitting \citep{chernozhukov2018double, van2018targeted, bickel1993efficient} is recommended. Here we describe one implementation of an augmented inverse probability weighted estimator based on the efficient influence function \citep[Theorem 2]{wen2023intervention}. We refer the reader to \citet{molina2017multiple}, \citet{rotnitzky2017multiply}, \citet{luedtke2017sequential}, and \citet{wen2022multiply} for a detailed review of various multiply robust algorithms, and to \citet{wen2023intervention} for multiply robust algorithms in the setting of non-deterministic treatment strategies. 

\begin{footnotesize}

\begin{enumerate}
\label{algorithm: AIPW}
    \item Build a data adaptive estimator of $f(z \mid l_1)$ and $f(a_{k} \mid \overline{l}_k, \overline{a}_{k-1}, y_{k-1} = 0, z)$.
    \item Set $\hat{h}_{K+1} = Y_K$.
    \item Set $q=0$ and let $k=K-q$.
    \begin{enumerate}
        \item Build a data adaptive estimator of $E[Y_{k} \mid \overline{L}_{k}, \overline{A}_{k}, \overline{Y}_{k-1} = 0, Z]$.
        \item If $Y_{k-1} = 1$, set $\hat{h}_{k} = 1$. Otherwise, compute 
        \begin{align*}
            \hat{h}_{k} = \begin{cases}
            \widehat{E}[\hat{h}_{k+1} \mid \overline{L}_{k}, A_k = 1, \overline{A}_{k-1}, Y_{k-1} = 0,Z] &\ : R_k = m, \\
            \pi_k^{inv}(\overline{L}_k, \overline{A}_{k-1}, Z) \times \widehat{E}[\hat{h}_{k+1} \mid \overline{L}_{k}, A_k = 1, \overline{A}_{k-1}, Y_{k-1} = 0, Z] \\
            + \left( 1- \pi_k^{inv}(\overline{L}_k, \overline{A}_{k-1}, Z) \right) \times \widehat{E}[\hat{h}_{k+1} \mid \overline{L}_{k}, A_k = 0, \overline{A}_{k-1}, Y_{k-1} = 0, Z] &\ : R_k < m \\
            \end{cases}
        \end{align*} 
        for a stochastic grace period treatment strategy where $\pi_k^{inv}$ may depend trivially on some of its arguments, or
        \begin{align*}
            \hat{h}_{k} = \begin{cases}
             \widehat{E}[\hat{h}_{k+1} \mid \overline{L}_{k}, A_k = 1, \overline{A}_{k-1}, Y_{k-1} = 0, Z] &\ : R_k = m, \\
             \widehat{E}[\hat{h}_{k+1} \mid \overline{L}_{k}, \overline{A}_{k}, Y_{k-1} = 0, Z] &\ : R_k < m
            \end{cases}
        \end{align*} 
        for a natural grace period treatment strategy. 
        \item Compute
         \begin{align*}
            \widehat{D}_{k} = \displaystyle \frac{\mathbbm{1}{(Z = z)}}{\widehat{f}(Z \mid L_1)} &\times \left\{ \prod_{j=1}^{k} \frac{f^{g_m}(A_{j} \mid \overline{L}_{j}, \overline{A}_{j-1}, y_{j-1} = 0, Z)}{\widehat{f}(A_{j} \mid \overline{L}_{j}, \overline{A}_{j-1}, y_{j-1} = 0, Z)} \right\} \\ 
            &\times \left( \hat{h}_{k+1} - \widehat{E}[\hat{h}_{k+1} \mid \overline{L}_{k}, \overline{A}_{k}, Y_{k-1} = 0, Z] \right) \times \left( 1-Y_{k-1} \right).
        \end{align*}
        \item If $k > 1$, then set $q = q+1$ and return to 3(a).
    \end{enumerate}
    \item Estimate the risk under the treatment strategy as:
    \begin{align*}
        \hat{\psi}^{AIPW} = \frac{1}{N} \sum_{i=1}^{N} \left( \hat{h}_{1,i} + \sum_{k=1}^{K} \widehat{D}_{k,i} \right).
    \end{align*}
    \item Estimate the variance, 
    \begin{align*}
        \widehat{Var}(\hat{\psi}^{AIPW}) = \frac{1}{N^2} \sum_{i=1}^{N} \left( \left( \hat{h}_{1,i} + \sum_{k=1}^{K} \widehat{D}_{k,i} \right) - \hat{\psi}^{AIPW} \right)^2.
    \end{align*}
\end{enumerate}

\end{footnotesize}

In Supplementary Materials Section~\ref{section: sample splitting} we describe how the algorithm can be modified to incorporate sample splitting.

For simplicity, thus far we have assumed no loss to follow-up. In Supplementary Materials Section~\ref{sec: extension to censoring} we extend the algorithm to accommodate censoring by loss to follow-up \citep{robins2000correcting}.

\section{Estimating the effect of antihypertensive medications under grace period strategies}
\label{section: data example}

We used KPC data to estimate the 3-year risk of acute myocardial infarction, congestive heart failure, ischemic stroke, or hemorrhagic stroke in individuals newly diagnosed with hypertension under various stochastic and natural grace period treatment strategies with initiation of either a thiazide diuretic or ACEI antihypertensive. Included individuals were aged 20 or older, diagnosed with hypertension (systolic blood pressure $\geq 140$, diastolic blood pressure $\geq 90$, or an ICD-9/ICD-10 hypertension diagnosis code), and initiated a first-line antihypertensive treatment with either a thiazide diuretic or ACEI prior to the year 2018. The variables included in $L_{k}$ are listed in Supplementary Materials Section~\ref{section: modelling details}. Individuals were not eligible if they had incomplete baseline information or a history of acute myocardial infarction, congestive heart failure, ischemic stroke, or hemorrhagic stroke prior to meeting the inclusion criteria. Individuals were followed until the end of 3 years or until loss to follow-up, defined as 24 months without a clinic encounter. KPC data contains information on pharmacy dispensings, and we defined $A_k$ as an indicator of having medication during month $k$. Under this definition, an individual has $A_k=1$ if they receive a medication dispensation in month $k$ or if they received a medication dispensation in a prior month that would be expected to last beyond the start of month $k$. This definition does not result in positivity violations for stochastic grace period regimes so long as $\overline{L}_{k}$ excludes the features of previous medication dispensations (i.e., their length). If these features must be included in $\overline{L}_{k}$ to satisfy the exchangeability condition discussed in Section~\ref{sec:identification}, to relax the positivity assumption, investigators might consider a stochastic regime that requires no intervention be made during months for which a past medication dispensation guarantees that $A_k=1$. This regime can be seen as a compromise between natural and stochastic grace period treatment strategies. 

We estimated the 3-year risk under interventions that initiated a thiazide diuretic and followed a natural grace period strategy with $m=3$, as well as stochastic grace period strategies with (1) $\pi^{inv}_k = 0.95$, (2) $\pi^{inv}_k = 0.90$, (3) $\pi^{inv}_k = 0.75$, and (4) $\pi^{inv}_k = 0.50$, each with $m=3$. We estimated risk under identical grace period strategies with initiation of an ACEI antihypertensive. In the Supplementary Materials, we considered strategies with $m=2$ and $m=4$.

To estimate risk under the treatment strategies of interest, we implemented the multiply robust algorithm detailed in \eqref{algorithm: AIPW}, extended to the setting with loss to follow-up (described in Supplementary Materials Section~\ref{sec: extension to censoring}). We estimated the conditional outcome and treatment probabilities in steps 1-4 of the algorithm using gradient boosted regression trees \citep{ridgeway2006gbm} as described further in Supplementary Materials Section~\ref{section: modelling details}.

\subsection{Results}

There were $27,182$ eligible individuals. The baseline characteristics for the cohort are summarized in Table~\ref{section:table 1}. Of those eligible, $67.8\%$ initiated treatment with an ACEI while 32.2$\%$ initiated treatment with a thiazide diuretic.
The 3-year risk estimates under each treatment strategy are presented in Table~\ref{data_example_results_m_3}, and 3-year risk differences are presented in Table~\ref{data_example_results_contrasts_m_3}.

\begin{singlespace}
\setlength{\tabcolsep}{1.5pt}
\begin{longtable}{ 
    l*{2}{S[round-precision=3]}
  }
\caption{3-year risk estimates and standard errors (s.e.) under various stochastic and natural grace period treatment strategies (each with m=3) following initiation of a thiazide or ACEI anti-hypertensive.} \label{data_example_results_m_3}
\\ 
\hline
\hline

Treatment strategy & {Estimate} & {s.e.}\\*

\toprule
\toprule
   
\ \ \uline{Initiate thiazide} \\*

Natural grace period strategy & 0.072 & 0.003\\*
Stochastic strategy with $\pi^{inv}_k = 0.95$ & 0.071 & 0.002\\*
Stochastic strategy with $\pi^{inv}_k = 0.90$ & 0.073 & 0.002\\*
Stochastic strategy with $\pi^{inv}_k = 0.75$ & 0.074 & 0.002\\*
Stochastic strategy with $\pi^{inv}_k = 0.50$ & 0.079 & 0.001\\*
%`Intention to treat' strategy & 0.074 & 0.002\\*

\ \ \uline{Initiate ACEI} \\*

Natural grace period strategy & 0.090 & 0.003\\*
Stochastic strategy with $\pi^{inv}_k = 0.95$ & 0.083 & 0.002\\*
Stochastic strategy with $\pi^{inv}_k = 0.90$ & 0.086 & 0.002\\*
Stochastic strategy with $\pi^{inv}_k = 0.75$ & 0.092 & 0.002\\*
Stochastic strategy with $\pi^{inv}_k = 0.50$ & 0.101 & 0.001\\*
%`Intention to treat' strategy & 0.097 & 0.002\\*

\bottomrule
\bottomrule
\end{longtable}
\end{singlespace}

\begin{singlespace}
\setlength{\tabcolsep}{1.5pt}
\begin{longtable}{ 
    l*{2}{S[round-precision=3]}
  }
\caption{3-year risk difference estimates and standard errors (s.e.) comparing initiation of a thiazide versus an ACEI anti-hypertensive under various stochastic and natural grace period treatment strategies (each with m=3).} \label{data_example_results_contrasts_m_3}
\\ 
\hline
\hline

Grace period strategy & {Estimate} & {s.e.}\\*

\toprule
\toprule
   
\ \ \uline{Initiate ACEI vs initiate thiazide} \\*

Natural grace period strategy & 0.018 & 0.004\\*
Stochastic strategy with $\pi^{inv}_k = 0.95$ & 0.012 & 0.003\\*
Stochastic strategy with $\pi^{inv}_k = 0.90$ & 0.013 & 0.003\\*
Stochastic strategy with $\pi^{inv}_k = 0.75$ & 0.019 & 0.003\\*
Stochastic strategy with $\pi^{inv}_k = 0.50$ & 0.022 & 0.002\\*
%`Intention to treat' strategy & 0.023 & 0.003\\*

\bottomrule
\bottomrule
\end{longtable}
\end{singlespace}

We estimated that the 3-year risk of the composite outcome was higher with stochastic grace period strategies that impose low medication adherence compared to those that impose high medication adherence. Specifically, for strategies that initiate a thiazide diuretic, the risk was $7.1\%$ ($95\%$ CI: $6.7$, $7.5\%$) under $95\%$ adherence compared to $7.9\%$ ($95\%$ CI: $7.7$, $8.1\%$) under $50\%$ adherence (risk difference: $-0.8$ percentage points, $95\%$ CI: $-1.2$, $-0.4$). This pattern was also observed for ACEI strategies, where the risk was $8.3\%$ ($95\%$ CI: $7.9$, $8.7\%$) under $95\%$ adherence compared to $10.1\%$ ($95\%$ CI: $9.9$, $10.3\%$) under $50\%$ adherence (risk difference: $-1.8$ percentage points, $95\%$ CI: $-2.2$, $-1.5$).

The estimated 3-year risk of the composite outcome was higher under a natural grace period strategy following initiation of ACEI compared to a thiazide diuretic agent ($9.0\% \text{ vs } 7.2\%$; risk difference: $1.8$ percentage points, $95\%$ CI: $1.0$, $2.6$). As shown in Table~\ref{data_example_results_contrasts_m_3}, the risk differences ranged from $1.2$ to $2.2$ percentage points for the comparison of thiziade versus ACEI stochastic grace period strategies with equivalent distributions of adherence. We estimated that the risk difference was largest when comparing thiazide to ACEI treatment strategies with $50\%$ medication adherence at each month within the grace period ($10.1\% \text{ vs } 7.9\%$; risk difference: $2.2$ percentage points, $95\%$ CI: $1.8$, $2.6$). 

The 3-year risk and risk difference estimates under treatment strategies with $m=2$ and $m=4$ are presented in Supplementary Materials Section~\ref{section:extra results}. As for treatment strategies with $m=3$, we estimated that risk was higher with strategies that impose low dispensing adherence than those that impose high medication adherence. We also estimated that risk differences varied by adherence, with the risk difference being largest when comparing thiazide to ACEI treatment strategies with $50\%$ medication adherence at each month within the grace period.

\section{Discussion}

When it is unrealistic to expect all individuals to take treatment continuously over follow-up, grace period regimes offer a pragmatic alternative for comparing the effectiveness of two or more treatments. The choice between stochastic and natural grace period strategies needs to be context specific, and each has advantages over the other. Because natural grace period strategies incorporate the natural adherence process, they may be more intuitive to follow in clinical practice. For example, individuals could be given a long-term prescription and instructed to dispense medication according to their natural inclinations, with their dispensations recorded so that any time they reach the end of a grace period a dispensation is forced, e.g., by delivery. For the same reason, the effectiveness of a natural grace period strategy depends not only on the `pharmacological' effectiveness of the initiated medication, but also on the natural utilization process in the target population. Because of this, risk estimates under natural grace period strategies are less transportable (to other populations) than estimates under stochastic strategies.

When two medications are compared, differences in their estimated effectiveness under natural grace period strategies might be due to differences in the natural adherence process of each medication, rather than their `pharmacological' effectiveness. This could arise when one medication is more likely to cause mild adverse effects that would not necessitate discontinuing the medication, but might lead to more frequent non-adherence. Similarly, differences in transient non-adherence may be observed when the logistics of obtaining treatment renewals (e.g., prescription refills, costs) are more challenging under one treatment than another. In such cases, natural grace period treatment strategy comparisons can be misleading when presented as differences in `pharmacological' effectiveness. 

Stochastic strategies, in contrast, can be specified such that the distribution of adherence is identical for both treatments under comparison. This specification can be useful when investigators are interested in a comparison of `pharmacological' effectiveness under less than continuous adherence because differences in outcomes between initiated medication groups could not be attributed to differential non-adherence. However, this requires that the stochastic strategy is not allowed to depend on covariate values that are themselves affected by the initiated medication.

We also showed that the identification conditions necessary for a causal interpretation of risk estimates under grace period treatment strategies depend not only on the causal relationship between variables, but also on strategy specific details. In particular, exchangeability assumptions might hold for some stochastic grace period treatment strategies, but not for natural grace period treatment strategies even when the causal relationship between the variables on the DAG are identical in both cases. Conversely, as discussed in Section~\ref{sec: gformula}, the positivity condition is guaranteed to hold for any observed history consistent with being within a grace period when a natural grace period treatment strategy is chosen, but not generally for stochastic grace period strategies. 

The results of our data example support recent findings suggesting that the risk of acute myocardial infarction, heart failure, and stroke is lower under first-line antihypertensive treatment with a thiazide diuretic compared to an ACEI \citep{suchard2019comprehensive}. Regardless of the grace period treatment strategy, we estimated that thiazide diuretic grace period treatment strategies reduced 3-year risk compared with ACEI strategies. Our data example also illustrates how risk differences can vary by adherence level, with risk differences being greatest when medication adherence is set to $50\%$ during grace periods. This suggests that risk differences estimated in prior studies may be specific to the natural adherence processes in those study populations. 

In the antihypertensive treatment setting, a range of choices could be reasonable for both $m$ and $\pi^{inv}_{k}$ since both short `drug holidays' and longer periods of non-utilization are common \citep{vrijens2008adherence}. In other settings where short `drug holidays' are uncommon and prolonged medication non-adherence is the norm, smaller choices of $m$ might result in the same positivity concerns inherent to the `per-protocol' estimand. Further, although they were not discussed in this paper, investigators might consider grace period estimands which allow the length of the grace period to vary over time. For example, this type of protocol might demand continuous adherence at the start of follow-up then allow more flexibility as follow-up progresses.

In studies comparing treatments using grace period strategies, we recommend that investigators consider providing estimates of risk under both stochastic grace period strategies with various levels of adherence and under natural grace period strategies.  

\section*{Acknowledgements}
We are grateful to Matthew F. Daley from Kaiser Permanente Colorado who assisted in data acquisition and commented on earlier versions of this manuscript.

\section*{Data availability statement}
The data were provided by Kaiser Permanente Colorado and cannot be shared publicly for the privacy of individuals that participated in the study. The code to reproduce the analysis is provided at \url{https://github.com/KerollosWanis/grace-periods}.

\section*{Funding statement}
This work was supported by NIH-NIDDK grant R01 DK 120598. 

\section*{Conflict of interest statement}
The authors have no conflicts of interest to declare. 

\bibliographystyle{rss}
\bibliography{bib.bib}

%%%%%%%%%% Merge with supplemental materials %%%%%%%%%%
\appendix
\begin{center}
\textbf{\large Supplemental Materials for ``Grace periods in comparative effectiveness studies of sustained treatments''}
\end{center}
%%%%%%%%%% Merge with supplemental materials %%%%%%%%%%
%%%%%%%%%% Prefix a "S" to all equations, figures, tables and reset the counter %%%%%%%%%%
\setcounter{equation}{0}
\setcounter{figure}{0}
\setcounter{table}{0}
\setcounter{page}{1}
\setcounter{section}{0}
\makeatletter
\renewcommand{\theequation}{S\arabic{equation}}
\renewcommand{\thefigure}{S\arabic{figure}}
\renewcommand{\thetable}{S\arabic{table}}
\renewcommand{\bibnumfmt}[1]{[S#1]}
\renewcommand{\thesection}{S\arabic{section}}
%%%%%%%%%% Prefix a "S" to all equations, figures, tables and reset the counter %%%%%%%%%%

\section{Examples of stochastic grace period strategies}
\label{section: stochastic grace period examples}
Example 1: Imagine a study where the investigator that assigns treatment as in \eqref{stochastic} with a two interval grace period ($m=2$) and with $\pi^{inv}_k(\overline{l}_k,\overline{a}_{k-1},z)$ depending only on $r_k$ such that

\begin{align}
   \pi^{inv}_k(r_k)= \begin{cases}
     1/3 &: r_k = 0 , \\
     1/2 &: r_k = 1 .
    \end{cases}
    \label{example: stochastic}
\end{align}

In words, under this strategy, in each interval $k$ and for an individual still surviving through $k-1$ who took treatment under the strategy at $k-1$ (i.e. they are in the first interval of a grace period), the investigator uses a random number generator to draw a standard uniform number for that individual, $X_k$, and, if $X_k \leq 1/3$, dispenses treatment in interval $k$, otherwise she does not dispense treatment. Alternatively, for a surviving individual who did not take treatment at $k-1$, if $X_k \leq 1/2$, the investigator dispenses treatment in that interval, otherwise she does not. As for any grace period strategy defined by $m=2$, if the individual has gone two intervals without treatment, the investigator dispenses treatment because the individual has reached the end of the specified grace period. In this example \eqref{example: stochastic} would induce a discrete uniform distribution of treatment over the grace period intervals among surviving individuals.

Example 2: Suppose $L_k$ is an indicator of a high risk medical condition in interval $k$. Now imagine a study where, again, the treatment strategy is a case of (\ref{stochastic}) with $m=2$ but now the investigator chooses $\pi^{inv}_k(\overline{l}_k,\overline{a}_{k-1},z)$ to depend on both $r_k$ and $l_k$ such that

\begin{align}
    \pi^{inv}_k(l_k, r_k)= \begin{cases}
     2/3 &: r_k = 0, l_k=1, \\
     1/3 &: r_k = 0, l_k=0, \\
     1/2 &: r_k = 1. \\
    \end{cases}
    \label{example: dynamic stochastic}
\end{align}

This strategy is identical to Example~\ref{example: stochastic}, except that individuals with $L_k=1$ are assigned twice the probability of treatment when they have taken treatment at $k-1$ compared to those with $L_k=0$. Unlike Example~\ref{example: stochastic}, this grace period strategy will not generally balance adherence for different initiated medications because $L_k$ may be affected by the choice of initiated medication.

In Supplementary Materials Section~\ref{section: grace period examples} we give examples of stochastic grace period strategies in clinical research.

\section{Constructing stochastic grace period treatment distributions}
\label{section:constructing_stochastic}

\begin{figure}
  \centering
    \includegraphics[width=1\textwidth]{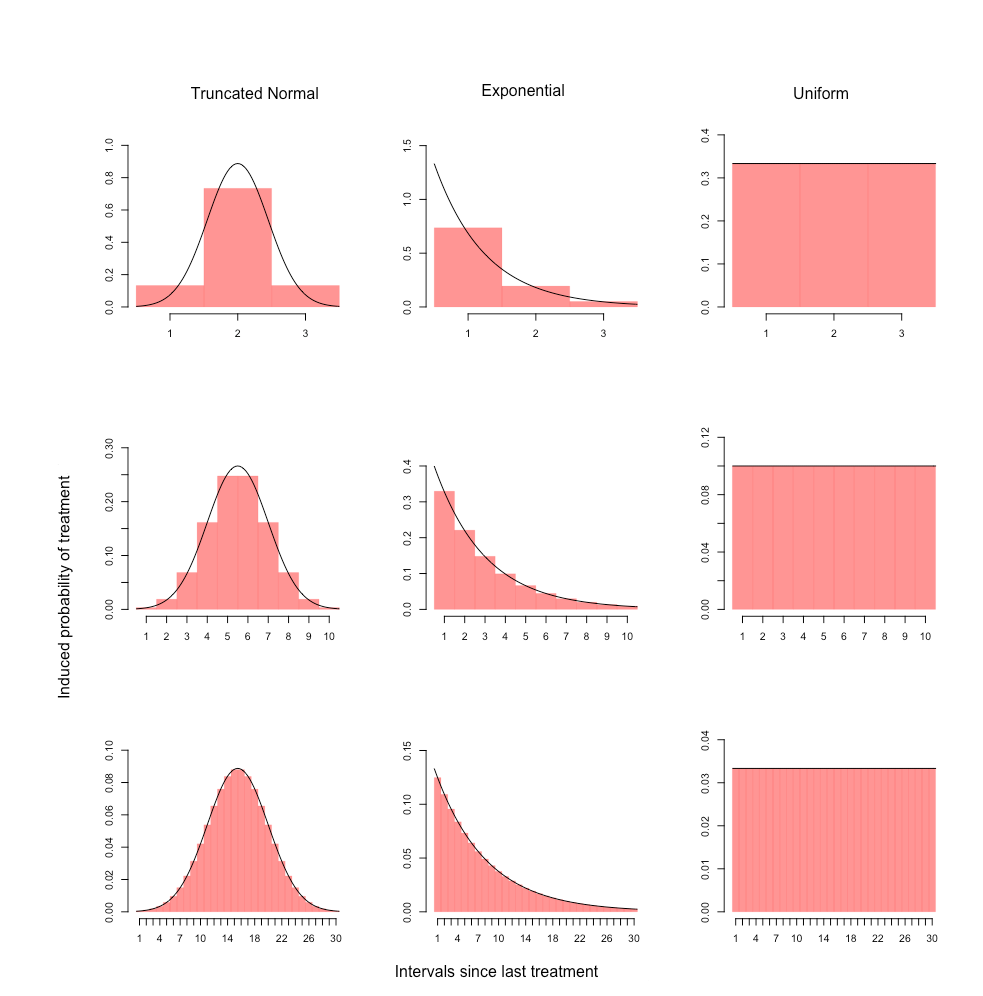}
  \caption{Distribution of treatment, $\Pr(A^{z,g_m+}_k=1, R_k=r_k)$, over a grace period with $m$ intervals under a stochastic grace period regime associated with a pre-specified probability density function $f_T$. Rows 1-3 show regimes with grace periods with m equal to 2, 9, and 29, respectively. Columns 1-3 show pre-specified probability density functions $f_T$, where $T$ follows a truncated normal distribution (on $(0,1)$, $\mu=0.5$, $\sigma=0.15$), a truncated exponential distribution (on $(0,1)$, $\lambda=4$), and a uniform distribution (on $(0,1)$), respectively.} \label{fig:joint_dists}
\end{figure}

\begin{figure}
  \centering
    \includegraphics[width=1\textwidth]{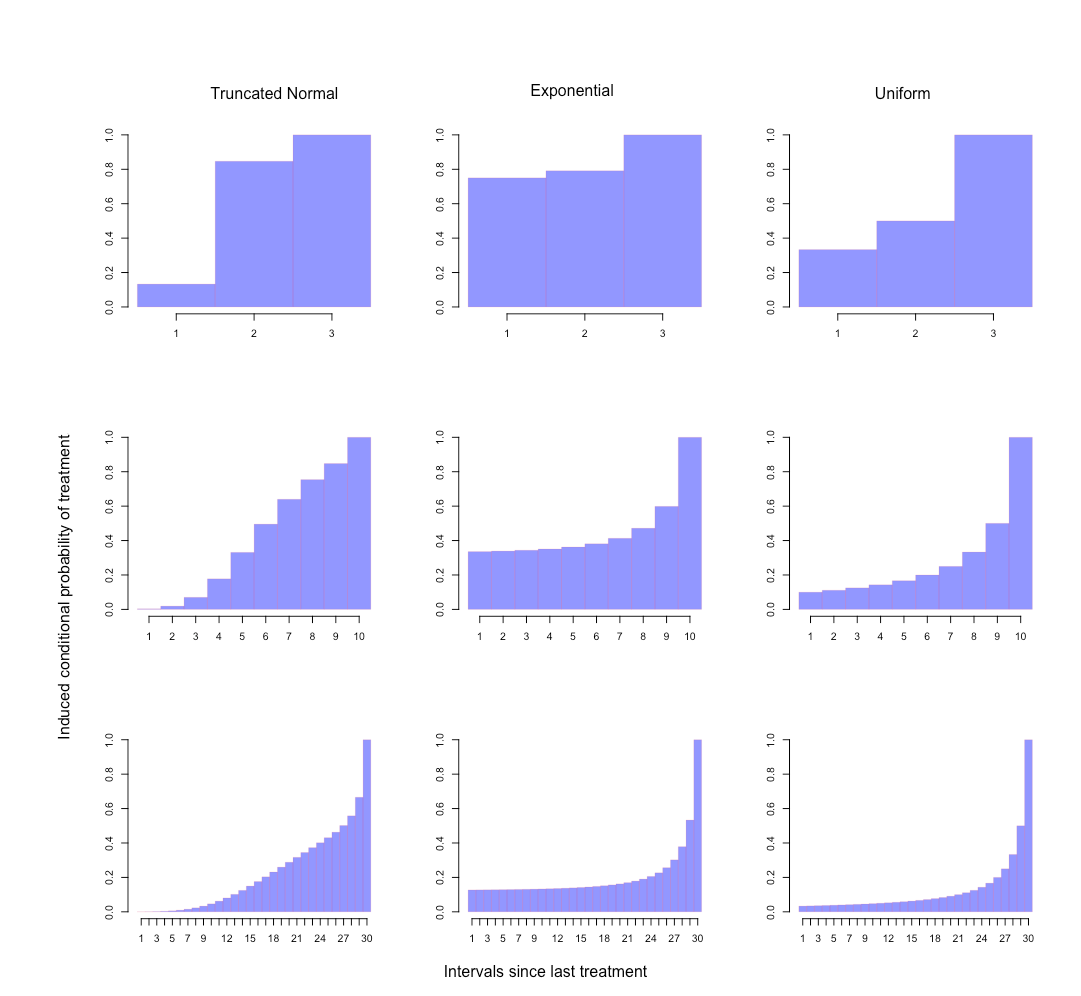}
  \caption{Conditional probabilities of treatment, $\pi_{k}^{inv}(r_k)$, over a grace period with $m$ intervals under a stochastic grace period regime associated with a pre-specified probability density function $f_T$. Rows 1-3 show regimes with grace periods with m equal to 2, 9, and 29, respectively. Columns 1-3 show pre-specified probability density functions $f_T$, where $T$ follows a truncated normal distribution (on $(0,1)$, $\mu=0.5$, $\sigma=0.15$), a truncated exponential distribution (on $(0,1)$, $\lambda=4$), and a uniform distribution (on $(0,1)$), respectively.} \label{fig:cond_dists}
\end{figure}

In some cases, the investigator may want to induce a discrete analogue of a particular continuous distribution over the grace period. For example, it may be desirable to have $\pi_{k}^{inv}(r_k)$ be a discretized truncated normal distribution or discrete uniform distribution (see Figure~\ref{fig:joint_dists} for illustrations). If we take $T$, with probability distribution function $f_T$ and support on the interval $(a, b)$, to be distributed according to the continuous distribution of interest, then we can induce a discretized analogue of $f_T$ by setting

\begin{align*}
    \Pr(A^{z,g_m+}_k=1, R_k=r_k) = \Pr\Bigg(\frac{(r_k+1) \times (b-a)}{m+1} > T-a \geq \frac{(r_k)\times(b-a)}{m+1}\Bigg)
\end{align*}

\noindent for $r_k = 0,\ldots,m$. The assigned probability of treatment given the number of intervals since last treatment is obtained by:

\begin{align*}
    \pi_{k}^{inv}(r_k) =
    \Pr(A^{z,g_m+}_k=1 | R_k=r_k) = 
    \frac{\Pr(A^{z,g_m+}_k=1, R_k=r_k)}{1-\sum_{j=0}^{r_k-1}\Pr(A^{z,g_m+}_k=1, R_k = j)}
\end{align*}
(see Figure~\ref{fig:cond_dists} for illustrations). Note that $\pi^{g_m}_k(r_k=m)$ is always equal to 1 since treatment must be taken when the number of intervals since last treatment equals the grace period length.

\pagebreak

\section{Examples of grace period strategies in clinical research}
\label{section: grace period examples}

\subsection{Natural grace period strategies}

\begin{itemize}
    \item Individuals with clinical depression are assigned cognitive behavioural therapy once every month, but they are allowed to skip their monthly session whenever they would prefer to, so long as they never skip more than $m=2$ consecutively. 
    \item Individuals with ulcerative colitis are prescribed infliximab to be taken every 8 weeks, but they are allowed to miss an infusion so long as they never miss more than $m=1$ in a row. 
    \item Individuals with a history of breast cancer are assigned to take tamoxifen daily for 5 years, except they are allowed to miss a pill on a given day so long as they never miss more than $m=7$ consecutively. 
    \item Individuals with type II diabetes are treated with metformin. Each time they go to the pharmacy to receive medication, they are given exactly 1 month's supply. They are to receive medication every month, but they are allowed to miss dispensations so long as they never miss more than $m=2$ in a row. 
    \item Individuals with hypertension are treated with a thiazide diuretic. They are given variable dispensation lengths - some receive 30 days of medication each time they go to the pharmacy, some 60 days, some 90 days, etc. They are to go to the pharmacy as often as they need to, such that they have medication each month, but they are allowed up to $m=3$ months in a row where they do not have any medication. That is, if they go 3 months in a row without having medication, they must go to the pharmacy and receive medication. 
    \item Individuals with hyperlipidemia are treated with a statin. They receive prescriptions for medication from their physician. They are to receive a prescription at the start of every 3-month study block, but they are allowed to miss getting a prescription so long as they do not miss more than $m=2$ prescriptions in a row. 
\end{itemize}

\subsection{Stochastic grace period strategies}

\begin{itemize}
    \item Individuals with clinical depression are assigned to receive cognitive behavioural therapy. The investigator flips a coin for each individual at the start of every month. If the coin comes up heads, the individual receives therapy. If tails, they do not receive therapy unless they have gone $m=2$ months in a row without getting therapy (because of prior coin flips) in which case the investigator ignores the coin and provides therapy. 
    \item Individuals with ulcerative colitis are prescribed infliximab. The investigator flips a coin for each individual at the start of each 8 week study block. If the coin comes up heads, the individual is administered an infusion. If tails, they are not administered an infusion unless they were not administered an infusion in the last study block ($m=1$) due to the prior coin flip, in which case the investigator ignores the coin and administers an infusion. 
    \item Individuals with a history of breast cancer are treated with tamoxifen for 5 years. Every day, they are to roll a 6-sided die. If the result of the roll is a 3 or higher, they take a pill. If the result of the roll is a 1 or 2, then they don't take a pill unless they have missed taking a pill for the past $m=7$ consecutive days (due to prior dice rolls), in which case they must take a pill. 
    \item Individuals with type II diabetes are treated with metformin. Each time they go to the pharmacy to receive medication, they are given exactly 1 month's supply. Every month, the investigator rolls a 4-sided die for each individual. If the result of the roll is a 2 or higher, then the individual is to go to their pharmacy and receive medication. If the result of the roll is a 1 then the individual is not to receive medication unless they have gone $m=2$ consecutive months without medication (due to prior dice rolls) in which case they must go receive medication. 
    \item Individuals with hypertension are treated with a thiazide diuretic. They are given variable dispensation lengths - some receive 30 days of medication each time they go to the pharmacy, some 60 days, some 90 days, etc. Whether they have medication in a given month depends on the roll of a 4-sided die. If the result of the roll is a 2 or higher, then they are to have medication in that month. If the result of the roll is a 1 then they are not to have medication in that month unless it has been $m=3$ consecutive months without having medication in which case they must receive medication. 
    \item Individuals with hyperlipidemia are treated with a statin. They receive prescriptions for medication from their physician. At the beginning of each 3-month study block the investigator flips a coin. If the result is heads, then the individual is given a prescription. If the result is tails then the individual is not given a prescription unless they have not been given prescriptions in both of the last $m=2$ study blocks (due to prior coin flips) in which case the investigator ignores the coin and provides a prescription. 
\end{itemize}

\pagebreak

\section{Simulation example}
 \label{sec: simulation example}
 
\subsection{Design}

We simulated data from a hypothetical randomized trial in which investigators compare two different drug treatments. Each individual is assigned to receive medication A ($Z=0$) or medication B ($Z=1$) for the duration of follow-up. Individuals are followed until death or the administrative end of the study (24 months following randomization). Individuals assigned to medication A cannot take medication B, and vice versa.

We generated data with $250,000$ individuals assigned to each treatment arm. We generated $A_k$ and $Y_k$ according to the following models:

\begin{itemize}
    \item $A_k$ was generated from a Bernoulli distribution with mean equal to $0.50 + (0.3 \times \mathbbm{1}{(R_k>0)}) + (0.3 \times \mathbbm{1}{(k=1)}) + (0.25 \times \mathbbm{1}{(Z=1)} \times \mathbbm{1}{(R_k=0)} \times \mathbbm{1}{(k>1)}) + (0.2 \times \mathbbm{1}{(R_k\geq3)})$. 
    \item $Y_k$ was generated from a Bernoulli distribution with mean equal to $0.03 + (-0.025 \times \mathbbm{1}{(A_k=1)})$.
\end{itemize}

In words, the data was generated with both initiated medications having identical effects on death in that the sharp null hypothesis holds for all individuals in the study population. Moreover, both medications reduce the risk of death when adhered to. We generated the data such that individuals who take either medication are less likely to take medication in the following interval, with the probability of not taking treatment being higher for those who took medication B. The data were generated for each individual, at each month, \textit{k}, from $k=1$ to $K=24$ or $Y_k=1$ (death), whichever came first.

\subsection{Estimation}

We estimated the counterfactual cumulative incidence curves under the following grace period treatment strategies, $z,g_{m, s}$ where $s$ indicates the intervention treatment distribution defined as follows:

\begin{itemize}
    \item $g_{0,1}$ is a static treatment strategy (alternatively a grace period strategy with $m=0$) which sets $A_k^{z,g_{0,1}+} = 1$ at every $k$;
    \item $g_{3,2}$ is a natural grace period treatment strategy with $m=3$;
    \item $g_{3,3}$ to $g_{3,8}$ are stochastic grace period treatment strategies defined by the following investigator specified treatment probability distributions, $\pi^{g_{3,3}}_k(r_k)$ to $\pi^{g_{3,8}}_k(r_k)$, each having $m=3$:
    \begin{footnotesize}
    \begin{multicols}{2}
    \begin{description}
        \item $\pi_k^{g_{3,3}}(r_k) = \begin{cases}
         0.9179567 &: r_k = 0, \\
         0.918423 &: r_k = 1, \\
         0.9241418 &: r_k = 2, \\
         1 &: r_k = 3;
        \end{cases}$
        \item $\pi_k^{g_{3,4}}(r_k) = \begin{cases}
         0.5552792 &: r_k = 0, \\
         0.5897977 &: r_k = 1, \\
         0.6791787 &: r_k = 2, \\
         1 &: r_k = 3;
        \end{cases}$
        \item $\pi_k^{g_{3,5}}(r_k) = \begin{cases}
         0.4063653 &: r_k = 0, \\
         0.6845376 &: r_k = 1, \\
         0.8639574 &: r_k = 2, \\
         1 &: r_k = 3;
        \end{cases}$
        \item $\pi_k^{g_{3,6}}(r_k) = \begin{cases}
         0.5817036 &: r_k = 0, \\
         0.8006517 &: r_k = 1, \\
         0.917605 &: r_k = 2, \\
         1 &: r_k = 3;
        \end{cases}$
        \item $\pi_k^{g_{3,7}}(r_k) = \begin{cases}
         0.75 &: r_k = 0, \\
         0.3\overline{3} &: r_k = 1, \\
         0.5 &: r_k = 2, \\
         1 &: r_k = 3;
        \end{cases}$
        \item $\pi_k^{g_{3,8}}(r_k) = \begin{cases}
         0.9 &: r_k = 0, \\
         0.3\overline{3} &: r_k = 1, \\
         0.5 &: r_k = 2, \\
         1 &: r_k = 3
        \end{cases}$
    \end{description}
    \end{multicols}
    \end{footnotesize}
\end{itemize}

where $\pi_k^{g_{3,3}}(r_k)$ and $\pi_k^{g_{3,4}}(r_k)$ are chosen such that the assigned distribution of treatment follows discrete analogues of truncated exponential distributions on $(0,1)$ with $\lambda=10$ and $\lambda=3$, respectively; $\pi_k^{g_{3,5}}(r_k)$ and $\pi_k^{g_{3,6}}(r_k)$ are chosen such that the assigned distribution of treatment follows discrete analogues of truncated normal distributions on $(0,1)$ with $\mu = 0.25, \sigma = 0.25$ and $\mu = 0.1, \sigma = 0.25$, respectively; and $\pi_k^{g_{3,7}}(r_k)$ and $\pi_k^{g_{3,8}}(r_k)$ are chosen such that the distribution of intervention treatment is discrete uniform after treatment is not taken for more than 1 interval.

In principle, a contrast between the counterfactual cumulative incidences under any of the above treatment strategies defines a causal effect. As such, causal effects may contrast treatment strategies which assign the same distribution of treatment while varying the assigned medication (e.g. $E[Y_k^{z=1, g_{0,1}}] - E[Y_k^{z=0, g_{0,1}}]$). But causal effects may also contrast treatment strategies which assign different distributions of treatment \textit{and} different medications; for example, the contrast $E[Y_k^{z=1, g_{3,4}}] - E[Y_k^{z=0, g_{3,3}}]$. Although not immediately obvious, and as described in the main text, the contrast $E[Y_k^{z=1, g_{3,2}}] - E[Y_k^{z=0, g_{3,2}}]$ which is defined by two natural grace period regimes, is an example of a contrast of two different medications under different treatment distributions when the natural treatment process differs by assigned medication. This is exactly the case for our data generating process where individuals assigned to medication B are less likely to take treatment than those assigned to medication A. Interpretation of causal effects which assign different distributions of treatment is challenging. Here we considered risk differences which, except in the case of natural grace period treatment strategies, compared assignment to medication A versus medication B under the same treatment distribution. 

Using correctly specified models, we estimated the counterfactual cumulative incidence curves under each of the above treatment strategies using inverse probability weighting, and computed the causal risk difference at the end of follow-up (24 months). We computed standard errors using a non-parametric bootstrap with 500 repetitions. 

\subsubsection{Results}

\begin{figure}
  \centering
    \includegraphics[width=1\textwidth]{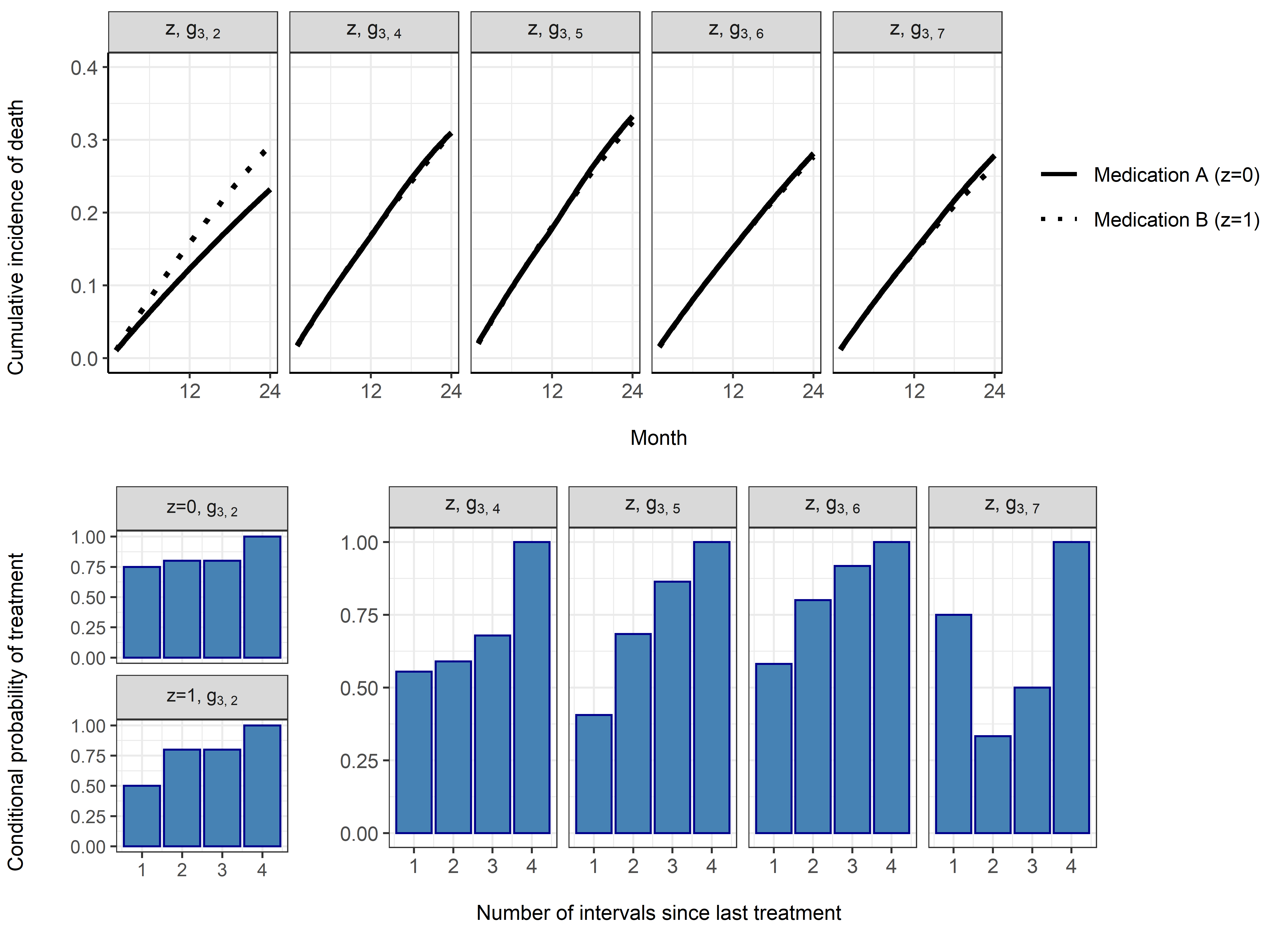}
  \caption{Cumulative incidence of death curves under various treatment strategies (top) and the conditional probability of treatment given the number of intervals since last treatment (bottom). $g_{3,2}$ is a natural grace period treatment strategy with \textit{m}=3, $g_{3,4}$, $g_{3,5}$, $g_{3,6}$, and $g_{3,7}$ are stochastic grace period treatment strategies.}
  \label{curves_and_hists}
\end{figure}

Figure \ref{curves_and_hists} shows the estimated counterfactual cumulative incidence curves for four of the treatment strategies, under assignment to medication A versus medication B. In the bottom panel, the conditional distribution of treatment for each of the treatment strategies is shown. For all of the stochastic grace period treatment strategies, the distribution of treatment is identical for assignment to medication A and medication B. In contrast, under a natural grace period treatment strategy, $g_{3,2}$, the distributions of treatment differ. 

\begin{figure}
  \centering
    \includegraphics[width=1\textwidth]{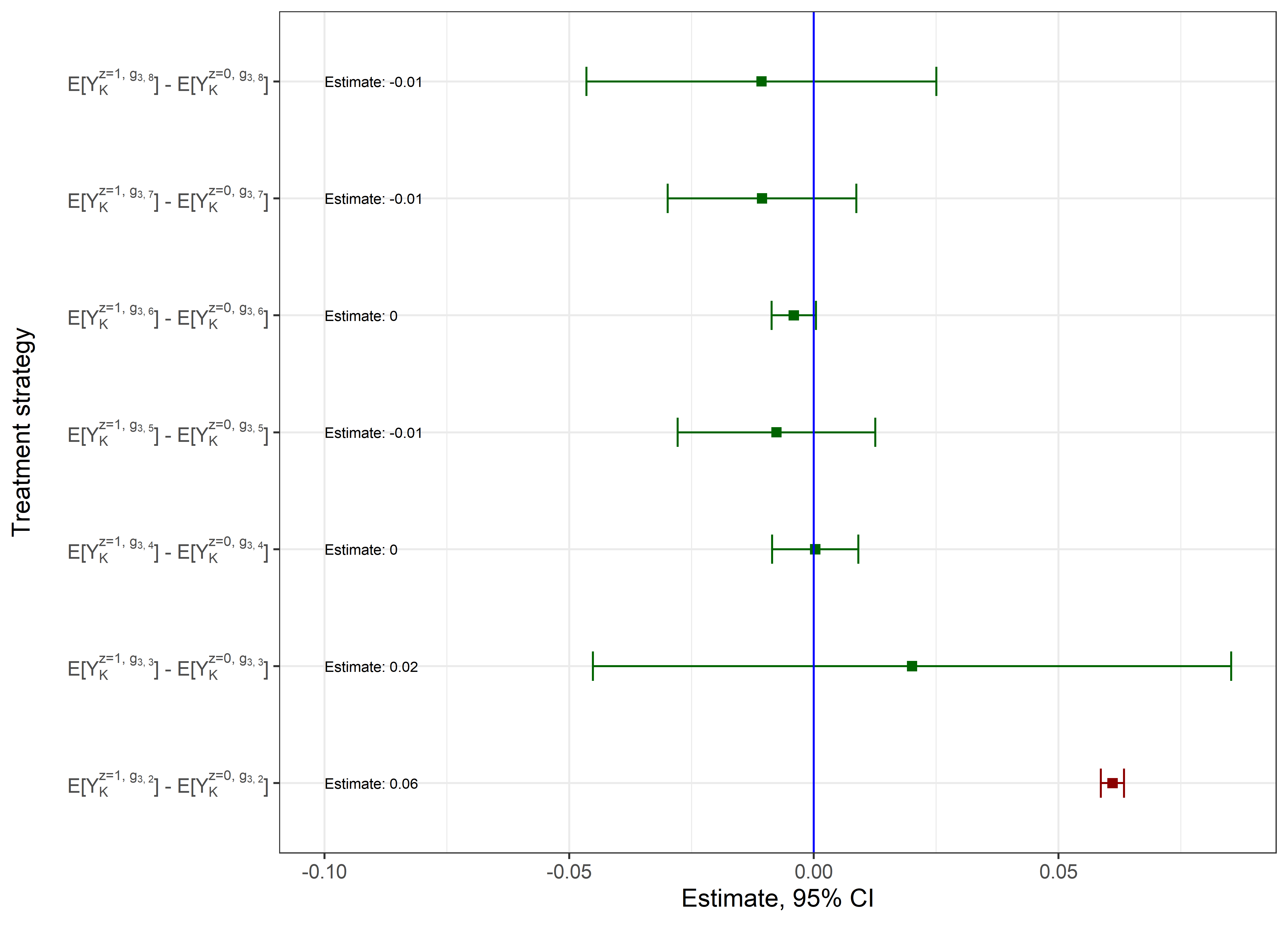}
  \caption{Estimates and 95$\%$ Wald confidence intervals of the causal risk difference at $K=24$ months. $g_{3,2}$ is a natural grace period treatment strategy with \textit{m}=3, $g_{3,3}$ to $g_{3,8}$ are stochastic grace period treatment strategies.}
  \label{causal_effect_by_protocol}
\end{figure}

Figure \ref{causal_effect_by_protocol} displays estimated causal risk differences under various intervention treatment distributions, comparing medication A to medication B. Estimated effects were close to null for all of the causal contrasts which involved stochastic grace period treatment strategies, but indicated a moderate protective effect of assignment to medication A versus medication B under a natural grace period treatment strategy. 

Despite the fact that we used a very large sample, estimated effects had wide confidence intervals when the intervention treatment distributions under stochastic grace period strategies deviated substantially from the true distribution of treatment. The most prominent example of this is for the static deterministic strategy, $g_{0,1}$, which was not estimable using the algorithm described in Section~\ref{algorithm: IPW} with correctly specified models since no individuals in our simulated sample (out of $250,000$) assigned to medication B took treatment in every interval and survived to the end of follow-up. A natural grace period treatment strategy, $g_{3,2}$, had the tightest confidence intervals. 

\pagebreak

\section{Single world intervention graphs}
\label{section: swig construction}

A SWIG specific to a particular grace period strategy $z,g_m$ can be constructed via a strategy (intervention) specific transformation of a causal DAG by: 1) splitting treatment nodes into two nodes, one indicating the natural value of treatment (i.e. $A_k^{z,g_m}$), and the other indicating the value of treatment under intervention (i.e. $A_k^{z,g_m+}$) with all arrows previously out of the treatment node now out of the node corresponding to intervention; 2) indexing variables affected by treatment using their counterfactual versions (e.g. $L_k^{z,g_m}$); and 3) when the treatment assignment rule depends on the values of other variables, adding dashed arrows from those variables into the intervention treatment nodes \citepS{richardson2013singleS}. To assess whether exchangeability for a particular grace period strategy $z, g_{m}$ is consistent with subject matter assumptions encoded in the causal DAG, familiar criteria depending on the ``absence of unblocked backdoor paths'' are applied to the resulting SWIG. 
 
\pagebreak

\section{Weighted representation of the g-formula}
\label{sec: IPW}

The g-formula \eqref{gformula}, characterized by a choice of $f^{g_m}(a_j \mid \overline{l}_j, \overline{a}_{j-1}, y_{j-1} = 0, z)$, can be re-expressed in several ways. For example, the following re-expression is an inverse probability weighted representation:

\begin{align}
    \sum_{k=1}^{K} \lambda_k \prod_{j=1}^{k-1} [1-\lambda_j] \label{eq: IPW_general1}
\end{align}

where

\begin{align}
    \lambda_k = \frac{E[Y_k(1-Y_{k-1})W_k]}{E[(1-Y_{k-1})W_k]} \label{eq: IPW_general2}
\end{align}

and weights, $W_k$, defined as

\begin{align}
  W_k = \displaystyle \frac{\mathbbm{1}{(Z = z)}}{f(Z \mid L_1)} \times \prod_{j=1}^k \frac{f^{g_m}(A_j \mid \overline{L}_j, \overline{A}_{j-1}, y_{j-1} = 0, Z)}{f(A_j \mid \overline{L}_j, \overline{A}_{j-1}, y_{j-1} = 0, Z)}. \label{eq: weights}
\end{align}

It follows from \eqref{fintdef} and \eqref{eq: f_int_natural_definition} that, for natural grace period treatment strategies, the numerator of \eqref{eq: weights} can be replaced by $f(A_j \mid \overline{L}_j, \overline{A}_{j-1}, y_{j-1} = 0, Z)$ whenever $R_j < m$. As such, for $R_j < m$, that is, in intervals prior to the end of the grace period, the j-specific contribution to $W_k$ evaluates to 1. But for $R_j = m$, the contribution evaluates to 0 if $A_j = 0$, and evaluates to $\frac{1}{\Pr[A_j = 1 \mid \overline{L}_j, \overline{A}_{j-1}, Y_{j-1} = 0, Z]}$ if $A_j = 1$. This allows us to re-write \eqref{eq: weights} as:

\begin{align}
  W_k = \displaystyle \frac{\mathbbm{1}{(Z = z)}}{f(Z \mid L_1)} \times \prod_{j=1}^k \frac{\mathbbm{1}{(R_{j+1} \leq m)}}{\Pr(R_{j+1} \leq m \mid \overline{L}_j, \overline{A}_{j-1}, Y_{j-1} = 0, Z)} \label{eq: weights_natural}
\end{align}

In Supplementary Materials Section~\ref{section:hypothetical_data_example_table}, we contrast the weights for a hypothetical data example under a natural grace period strategy, a simplified stochastic grace period strategy, and a static `always treat' strategy. 

If $f(z \mid l_1)$ and $f(a_{k} \mid \overline{l}_k, \overline{a}_{k-1}, y_{k-1} = 0,z)$ can be modelled correctly using parametric methods, then a simple inverse probability weighted estimator will be consistent, and valid confidence intervals can be obtained using standard methods \citepS{hernan2020causalS}. The algorithm proceeds as follows:

\begin{footnotesize}

\begin{enumerate}
\label{algorithm: IPW}
    \item Fit a parametric regression model for $f(z \mid l_1)$ and $f(a_{k} \mid \overline{l}_k, \overline{a}_{k-1}, y_{k-1} = 0,z)$, the conditional treatment probabilities. For example, we might assume a pooled logistic regression model. 
    \item Compute        
         \begin{align*}
            \widehat{W}_{k} = \displaystyle \frac{\mathbbm{1}{(Z = z)}}{\widehat{f}(Z \mid L_1)} \times \left\{ \prod_{j=1}^{k}\frac{ f^{g_m}(A_{j} \mid \overline{L}_{j}, \overline{A}_{j-1}, y_{j-1} = 0, Z)}{\widehat{f}(A_{j} \mid \overline{L}_{j}, \overline{A}_{j-1}, y_{j-1} = 0, Z)} \right\}
        \end{align*}
    \item Compute the risk of failure by the end of $K$ as $\sum_{k=1}^{K} \widehat{\lambda}_k \displaystyle  \prod_{j=1}^{k-1} [1-\widehat{\lambda}_j]$ with the estimated weights, $\widehat{W}_k$, in $\widehat{\lambda}_k = \frac{\widehat{E}[Y_k(1-Y_{k-1})\widehat{W}_k]}{\widehat{E}[(1-Y_{k-1})\widehat{W}_k]}$.
\end{enumerate}

\end{footnotesize}

\pagebreak

\section{Hypothetical data example}
\label{section:hypothetical_data_example_table}

\begin{singlespace}
\begin{tiny}
\setlength{\tabcolsep}{5pt}
\begin{longtable}{ 
    P{1cm}P{1cm}P{1cm}P{1.5cm}P{3cm}P{3cm}P{3cm}
  }
\caption{ Data for four hypothetical individuals whose treatment histories are all consistent with natural and stochastic grace period treatment strategies with $m = 2$ the grace period length. The stochastic grace period treatment strategy sets $\pi^{inv}_k(r_k) = 0.8$ for $r_k < m$ at every interval $k$. $p_{i,k}$ is the observed probability that individual $i$ takes treatment at time $k$ given their past treatment and covariate history. We contrast these weights with those from a `always treat' strategy (i.e. a grace period strategy with $m=0$)}\\ 
  \hline
  \hline
Person ($i$) & Time ($k$) & Treatment ($A_k$) & Number of intervals since last treatment ($R_k$) & $W_k$ under always treat strategy, $m=0$ & $W_k$ under natural grace period treatment strategy & $W_k$ under stochastic grace period treatment strategy \\
   \toprule
   \toprule
1 & 1 & 1 & 0 & $1/p_{1,1}$ & 1 & $0.8/p_{1,1}$ \\
1 & 2 & 1 & 0 & $1/[p_{1,1} \times p_{1,2}]$ & 1 & $0.8^2/[p_{1,1} \times p_{1,2}]$ \\
1 & 3 & 1 & 0 & $1/[p_{1,1} \times p_{1,2} \times p_{1,3}]$ & 1 & $0.8^3/[p_{1,1} \times p_{1,2} \times p_{1,3}]$ \\
1 & 4 & 1 & 0 & $1/[p_{1,1} \times p_{1,2} \times p_{1,3} \times  p_{1,4}]$ & 1 & $0.8^4/[p_{1,1} \times p_{1,2} \times p_{1,3} \times p_{1,4}]$ \\
1 & 5 & 1 & 0 & $1/[p_{1,1} \times p_{1,2} \times p_{1,3} \times  p_{1,4} \times p_{1,5}]$ & 1 & $0.8^5/[p_{1,1} \times p_{1,2} \times p_{1,3} \times p_{1,4} \times p_{1,5}]$  \\
\hdashline
2 & 1 & 1 & 0 & $1/p_{2,1}$ & 1 & $0.8/p_{2,1}$ \\
2 & 2 & 0 & 0 & 0 & 1 & $(0.8 \times 0.2)/[p_{2,1} \times (1-p_{2,2})]$ \\
2 & 3 & 1 & 1 & 0 & 1 & $(0.8^2 \times 0.2)/[p_{2,1} \times (1-p_{2,2}) \times p_{2,3}]$ \\
2 & 4 & 0 & 0 & 0 & 1 & $(0.8^2 \times 0.2^2)/[p_{2,1} \times (1-p_{2,2}) \times p_{2,3} \times (1-p_{2,4})]$ \\
2 & 5 & 1 & 1 & 0 & 1 & $(0.8^3 \times 0.2^2)/[p_{2,1} \times (1-p_{2,2}) \times p_{2,3} \times (1-p_{2,4}) \times p_{2,5}]$ \\
\hdashline
3 & 1 & 1 & 0 & $1/p_{3,1}$ & 1 & $0.8/p_{3,1}$ \\
3 & 2 & 0 & 0 & 0 & 1 & $(0.8 \times 0.2)/[p_{3,1} \times (1-p_{3,2})]$ \\
3 & 3 & 0 & 1 & 0 & 1 & $(0.8 \times 0.2^2)/[p_{3,1} \times (1-p_{3,2}) \times (1-p_{3,3})]$ \\
3 & 4 & 0 & 2 & 0 & 0 & 0 \\
3 & 5 & 1 & 3 & 0 & 0 & 0 \\
\hdashline
4 & 1 & 1 & 0 & $1/p_{4,1}$ & 1 & $0.8/p_{4,1}$ \\
4 & 2 & 0 & 0 & 0 & 1 & $(0.8 \times 0.2)/[p_{4,1} \times (1-p_{4,2})]$ \\
4 & 3 & 0 & 1 & 0 & 1 & $(0.8 \times 0.2^2)/[p_{4,1} \times (1-p_{4,2}) \times (1-p_{4,3})]$ \\
4 & 4 & 1 & 2 & 0 & $1/p_{4,4}$ & $(0.8^2 \times 0.2^2)/[p_{4,1} \times (1-p_{4,2}) \times (1-p_{4,3}) \times p_{4,4}]$ \\
4 & 5 & 0 & 0 & 0 & $1/p_{4,4}$ & $(0.8^2 \times 0.2^3)/[p_{4,1} \times (1-p_{4,2}) \times (1-p_{4,3}) \times p_{4,4} \times (1-p_{4,5})]$ \\
\bottomrule
\bottomrule
\label{tab:hypothetical_data_example}
\end{longtable}
\end{tiny}
\end{singlespace}

\pagebreak

\section{Incorporating sample splitting}
\label{section: sample splitting}

The algorithm in Section~\ref{sec: AIPW} can be modified to incorporate sample splitting by randomly splitting the observations, denoted $i$ in $1, \ldots, N$, into $P$ disjoint partitions. We let $p_i$ indicate the partition membership, $p_i \in 1,\ldots,P$, such that, for example, an individual observation randomly assigned to the $\zeta$ partition would have $p_i = \zeta$. Let $\boldsymbol{\mathbbm{D}}_{tr} = \{ (\overline{L}_{K,i}, \overline{A}_{K,i}, \overline{Y}_{K,i}): p_i \neq \zeta \}$ be the `training' subset of observations, $\boldsymbol{\mathbbm{D}}_{est} = \{ (\overline{L}_{K,i}, \overline{A}_{K,i}, \overline{Y}_{K,i}): p_i = \zeta \}$ be the `estimation' subset of observations, and $\boldsymbol{\mathbbm{D}} = \boldsymbol{\mathbbm{D}}_{tr} \cup \boldsymbol{\mathbbm{D}}_{est}$ be all the full set of observations. The algorithm proceeds as detailed above, except that each step where a data adaptive estimator is fit, the estimator is fit (`trained') only using $\boldsymbol{\mathbbm{D}}_{tr}$ while predicted values are obtained in the full set of observations, $\boldsymbol{\mathbbm{D}}$. At step 5 and 6 of the algorithm, $\hat{\psi}^{AIPW}_{\zeta}$ and $\widehat{Var}(\hat{\psi}^{AIPW}_{\zeta})$ are computed using only observations from partition $\zeta$. The algorithm is repeated for each $\zeta \in 1,\ldots, P$. The final estimate $\hat{\psi}^{AIPW}$ is obtained by averaging over the $P$ estimates, and the final estimate of the variance is given by $\widehat{Var}(\hat{\psi}^{AIPW}) = \frac{1}{P^2} \left( \widehat{Var}(\hat{\psi}^{AIPW}_{1}) + \ldots + \widehat{Var}(\hat{\psi}^{AIPW}_{P})) \right)$.

\pagebreak

\section{Generalization of the multiply robust algorithm to allow for censoring due to loss to follow-up}
\label{sec: extension to censoring}

In the data example presented in the main text we estimated the risk of failure for eligible individuals had they initiated treatment with an anti-hypertensive medication, had they followed natural or stochastic grace period adherence strategies, \textit{and} had censoring due to loss to follow-up been abolished. In this section we describe how the multiply robust algorithm in Section~\ref{sec: AIPW} can be modified to estimate the effect of grace period treatment strategies with censoring abolished. The sufficient conditions that allow identification of the risk of failure under a strategy which abolishes loss to follow-up are discussed in detail elsewhere \citepS{hernan2020causalS}.

Define $C_k$ as an indicator of loss to follow-up before the end of interval $k$, and make the temporal ordering assumption $(L_{k},A_{k},C_{k},Y_{k})$. Then the following algorithm estimates the risk of failure by time $K$ under a natural or stochastic grace period strategy which also abolishes censoring due to loss to follow-up. 

\begin{footnotesize}

\begin{enumerate}
    \item Build a data adaptive estimator of $f(z \mid l_1)$, $f(a_{k} \mid \overline{l}_k, \overline{a}_{k-1}, c_{k-1} = y_{k-1} = 0, z)$, and $f(c_{k} \mid \overline{l}_k, \overline{a}_{k}, c_{k-1} = y_{k-1} = 0, z)$.
    \item Set $\hat{h}_{K+1} = Y_K$.
    \item Set $q=0$ and let $k=K-q$.
    \begin{enumerate}
        \item Build a data adaptive estimator of $E[Y_{k} \mid \overline{L}_{k}, \overline{A}_{k}, C_{k} = Y_{k-1} = 0, Z]$.
        \item If $Y_{k-1} = 1$, set $\hat{h}_{k} = 1$. Otherwise, compute  
        \begin{align*}
            \hat{h}_{k} = \begin{cases}
            \widehat{E}[\hat{h}_{k+1} \mid \overline{L}_{k}, A_k = 1, \overline{A}_{k-1}, C_{k} = Y_{k-1} = 0, Z] &\ : R_k = m, \\
            \pi_k^{inv}(\overline{L}_k, \overline{A}_{k-1}, Z) \times \widehat{E}[\hat{h}_{K+1} \mid \overline{L}_{k}, A_k = 1, \overline{A}_{k-1}, C_{k} = Y_{k-1} = 0, Z] \\
            + \left( 1- \pi_k^{inv}(\overline{L}_k, \overline{A}_{k-1}, Z) \right) \times \widehat{E}[\hat{h}_{k+1} \mid \overline{L}_{k}, A_k = 0, \overline{A}_{k-1}, C_{k} = Y_{k-1} = 0, Z] &\ : R_k < m \\
            \end{cases}
        \end{align*} 
        for a stochastic grace period treatment strategy where $\pi_k^{inv}$ may depend trivially on some of its arguments, or
        \begin{align*}
            \hat{h}_{k}^{z,g_m} = \begin{cases}
             \widehat{E}[\hat{h}_{k+1} \mid \overline{L}_{k}, A_k = 1, \overline{A}_{k-1}, C_{k} = Y_{k-1} = \overline{0}, Z] &\ : R_k = m, \\
             \widehat{E}[\hat{h}_{k+1} \mid \overline{L}_{k}, \overline{A}_{k}, C_{k} = Y_{k-1} = 0, Z] &\ : R_k < m
            \end{cases}
        \end{align*} 
        for a natural grace period treatment strategy. 
        \item Compute 
         \begin{align*}
            \widehat{D}_{k} = \displaystyle \frac{\mathbbm{1}{(Z = z)}}{\widehat{f}(Z \mid L_1)} \times \Bigg\{ &\prod_{j=1}^{k} \frac{f^{g_m}(A_{j} \mid \overline{L}_{j}, \overline{A}_{j-1}, c_{j-1} = y_{j-1} = 0, Z)}{\widehat{f}(A_{j} \mid \overline{L}_{j}, \overline{A}_{j-1}, c_{j-1} = y_{j-1} = 0, Z)} \\
            & \times \prod_{j=1}^{k} \frac{\mathbbm{1}{(C_{j} = 0)}}{\widehat{f}(C_{j} \mid \overline{L}_{j}, \overline{A}_{j}, c_{j-1} = y_{j-1} = 0, Z)} \Bigg\} \\ 
            \times &\left( \hat{h}_{K+1} - \widehat{E}[\hat{h}_{K+1} \mid \overline{L}_{K}, \overline{A}_{K}, C_{K} = Y_{K-1} = 0, Z] \right) \times \left( 1-Y_{k-1} \right).
        \end{align*}
    \item If $k > 1$, then set $q = q+1$ and return to 3(a).
    \end{enumerate}
    \item Compute the expected risk under the treatment strategy as:
    \begin{align*}
        \hat{\psi}^{AIPW} = \frac{1}{N} \sum_{i=1}^{N} \left( \hat{h}_{1,i} + \sum_{k=1}^{K} \widehat{D}_{k,i} \right)
    \end{align*}
    \item Compute the variance, 
    \begin{align*}
        \widehat{Var}(\hat{\psi}^{AIPW}) = \frac{1}{N^2} \sum_{i=1}^{N} \left( \left( \hat{h}_{1,i} + \sum_{k=1}^{K} \widehat{D}_{k,i} \right) - \hat{\psi}^{AIPW} \right)^2
    \end{align*}
\end{enumerate}
\end{footnotesize}

As detailed in the Supplementary Materials Section~\ref{section: sample splitting}, this algorithm can be modified to incorporate sample splitting.

\pagebreak

\section{Data example: nuisance parameter estimation}
\label{section: modelling details}

We estimated the effect of various stochastic and natural grace period treatment strategies with initiation of either a thiazide or ACEI anti-hypertensive on the 3-year risk of acute myocardial infarction, congestive heart failure, ischemic stroke, or hemorrhagic stroke in individuals newly diagnosed with hypertension using the algorithm described in Section~\ref{algorithm: AIPW}. We estimated the conditional (on treatment, $A_k$, and confounder, $L_k$, histories) outcome and treatment probabilities using gradient boosted regression trees \citepS{ridgeway2006gbmS}, selecting hyperparameters with a grid search over the following ranges: shrinkage between 0.01 and 0.1 in increments of 0.01, number of trees between 100 and 500 in increments of 50, and maximum tree depth between 1 and 5 in increments of 1. The estimation algorithm included confounder histories by adding the lagged values of variables to the gradient boosted tree input dataset up to the 5\textsuperscript{th} lagged value. An \textit{i}\textsuperscript{th} lagged value of a variable relative to $k$ is its value at $k-i$. The hyperparameter combinations which minimized mean squared error (for each iterative outcome model separately) or cross-validation binary log-loss (for treatment models) using 5-fold cross validation were chosen. Weights were truncated to the 99\textsuperscript{th} percentile \citepS{cole2008constructingS}. $L_k$ included indicators of the following diagnoses by month $k$: bariatric surgery, chronic kidney disease, liver cirrhosis, type 2 diabetes, and type 1 diabetes, $L_1$ additionally included the following time-fixed characteristics: age at baseline, sex, and any pre-baseline history of the variables in $L_k$. 

\pagebreak

\section{Summary of the baseline characteristics for the cohort of eligible individuals from the Kaiser Permanente Colorado cohort}
\label{section:table 1}

\begin{longtable}{lrr}
  \hline
 & Initiated ACEI & Initiated thiazide \\ 
  \hline
N & 18,429 & 8,753 \\ 
  Female gender, n (\%) & 8,203 (44.5) & 5,726 (65.4) \\ 
  Bariatric diagnosis, n (\%) & 49 (0.3) & 19 (0.2) \\ 
  Chronic kidney disease, n (\%) & 1,749 (9.5) & 371 (4.2) \\ 
  Liver Cirrhosis, n (\%) & 42 (0.2) & 27 (0.3) \\ 
  Type 2 diabetes, n (\%) & 5,112 (27.7) & 723 (8.3) \\ 
  Type 1 diabetes, n (\%) & 269 (1.5) & 28 (0.3) \\ 
  Age (years), median [IQR] & 58.7 [49.5, 67.0] & 57.3 [47.5, 66.3] \\ 
   \hline
\hline
\end{longtable}

\pagebreak

\section{3-year risk and risk difference estimates under stochastic and natural grace period treatment strategies with m=2 and m=4}
\label{section:extra results}

\begin{singlespace}
\setlength{\tabcolsep}{1.5pt}
\begin{longtable}{ 
    l*{2}{S[round-precision=3]}
  }
\caption{3-year risk estimates and standard errors (s.e.) under various stochastic and natural grace period treatment strategies (each with m=2) following initiation of a thiazide or ACEI anti-hypertensive.} \label{data_example_results_m_2}
\\ 
\hline
\hline

Treatment strategy & {Estimate} & {s.e.}\\*

\toprule
\toprule
   
\ \ \uline{Initiate thiazide} \\*

Natural grace period strategy & 0.068 & 0.003\\*
Stochastic strategy with $\pi^{inv}_k = 0.95$ & 0.068 & 0.002\\*
Stochastic strategy with $\pi^{inv}_k = 0.90$ & 0.070 & 0.002\\*
Stochastic strategy with $\pi^{inv}_k = 0.75$ & 0.071 & 0.002\\*
Stochastic strategy with $\pi^{inv}_k = 0.50$ & 0.077 & 0.001\\*

\ \ \uline{Initiate ACEI} \\*

Natural grace period strategy & 0.083 & 0.003\\*
Stochastic strategy with $\pi^{inv}_k = 0.95$ & 0.083 & 0.002\\*
Stochastic strategy with $\pi^{inv}_k = 0.90$ & 0.085 & 0.002\\*
Stochastic strategy with $\pi^{inv}_k = 0.75$ & 0.089 & 0.002\\*
Stochastic strategy with $\pi^{inv}_k = 0.50$ & 0.097 & 0.001\\*

\bottomrule
\bottomrule
\end{longtable}
\end{singlespace}

\begin{singlespace}
\setlength{\tabcolsep}{1.5pt}
\begin{longtable}{ 
    l*{2}{S[round-precision=3]}
  }
\caption{3-year risk difference estimates and standard errors (s.e.) comparing initiation of a thiazide versus an ACEI anti-hypertensive under various stochastic and natural grace period treatment strategies (each with m=2).} \label{data_example_results_contrasts_m_2}
\\ 
\hline
\hline

Grace period strategy & {Estimate} & {s.e.}\\*

\toprule
\toprule
   
\ \ \uline{Initiate ACEI vs initiate thiazide} \\*

Natural grace period strategy & 0.015 & 0.004\\*
Stochastic strategy with $\pi^{inv}_k = 0.95$ & 0.015 & 0.003\\*
Stochastic strategy with $\pi^{inv}_k = 0.90$ & 0.014 & 0.003\\*
Stochastic strategy with $\pi^{inv}_k = 0.75$ & 0.020 & 0.003\\*
Stochastic strategy with $\pi^{inv}_k = 0.50$ & 0.020 & 0.002\\*

\bottomrule
\bottomrule
\end{longtable}
\end{singlespace}

\pagebreak

\begin{singlespace}
\setlength{\tabcolsep}{1.5pt}
\begin{longtable}{ 
    l*{2}{S[round-precision=3]}
  }
\caption{3-year risk estimates and standard errors (s.e.) under various stochastic and natural grace period treatment strategies (each with m=4) following initiation of a thiazide or ACEI anti-hypertensive.} \label{data_example_results_m_4}
\\ 
\hline
\hline

Treatment strategy & {Estimate} & {s.e.}\\*

\toprule
\toprule
   
\ \ \uline{Initiate thiazide} \\*

Natural grace period strategy & 0.076 & 0.003\\*
Stochastic strategy with $\pi^{inv}_k = 0.95$ & 0.072 & 0.002\\*
Stochastic strategy with $\pi^{inv}_k = 0.90$ & 0.073 & 0.002\\*
Stochastic strategy with $\pi^{inv}_k = 0.75$ & 0.077 & 0.002\\*
Stochastic strategy with $\pi^{inv}_k = 0.50$ & 0.079 & 0.001\\*

\ \ \uline{Initiate ACEI} \\*

Natural grace period strategy & 0.089 & 0.002\\*
Stochastic strategy with $\pi^{inv}_k = 0.95$ & 0.084 & 0.002\\*
Stochastic strategy with $\pi^{inv}_k = 0.90$ & 0.086 & 0.002\\*
Stochastic strategy with $\pi^{inv}_k = 0.75$ & 0.091 & 0.002\\*
Stochastic strategy with $\pi^{inv}_k = 0.50$ & 0.101 & 0.002\\*

\bottomrule
\bottomrule
\end{longtable}
\end{singlespace}

\begin{singlespace}
\setlength{\tabcolsep}{1.5pt}
\begin{longtable}{ 
    l*{2}{S[round-precision=3]}
  }
\caption{3-year risk difference estimates and standard errors (s.e.) comparing initiation of a thiazide versus an ACEI anti-hypertensive under various stochastic and natural grace period treatment strategies (each with m=4).} \label{data_example_results_contrasts_m_4}
\\ 
\hline
\hline

Grace period strategy & {Estimate} & {s.e.}\\*

\toprule
\toprule
   
\ \ \uline{Initiate ACEI vs initiate thiazide} \\*

Natural grace period strategy & 0.013 & 0.003\\*
Stochastic strategy with $\pi^{inv}_k = 0.95$ & 0.012 & 0.003\\*
Stochastic strategy with $\pi^{inv}_k = 0.90$ & 0.013 & 0.003\\*
Stochastic strategy with $\pi^{inv}_k = 0.75$ & 0.015 & 0.003\\*
Stochastic strategy with $\pi^{inv}_k = 0.50$ & 0.022 & 0.002\\*

\bottomrule
\bottomrule
\end{longtable}
\end{singlespace}

\pagebreak

\bibliographystyleS{rss}
\bibliographyS{bibAppendix.bib}

\end{document}